\documentclass[12pt]{article}
\usepackage{amsfonts,amssymb,epsfig,amsmath,mathtools}
\usepackage{color}

\renewcommand{\baselinestretch}{1.2}

\newcommand{\be}{\begin{eqnarray}}
\newcommand{\ee}{\end{eqnarray}}

\newcommand{\bn}{\begin{enumerate}}
\newcommand{\en}{\end{enumerate}}

\begin{document}

\makeatletter \@addtoreset{equation}{section} \makeatother
\renewcommand{\theequation}{\thesection.\arabic{equation}}
\renewcommand{\thefootnote}{\alph{footnote}}

\begin{titlepage}

\begin{center}

\hfill {\tt SNUTP19-001}\\

\vspace{3cm}

{\Large\bf Large AdS$_6$ black holes from CFT$_5$}

\vspace{2cm}

\renewcommand{\thefootnote}{\alph{footnote}}

{\large Sunjin Choi$^{1,2}$ and Seok Kim$^1$}

\vspace{0.7cm}

\textit{$^1$Department of Physics and Astronomy \& Center for
Theoretical Physics,\\
Seoul National University, 1 Gwanak-ro, Gwanak-gu, Seoul 08826, Republic of Korea}\\

\vspace{0.2cm}

\textit{$^2$School of Physics, Korea Institute for Advanced Study,\\
85 Hoegi-ro, Dongdaemun-gu, Seoul 02455, Republic of Korea}

\vspace{0.7cm}

E-mails: {\tt sunjinchoi@kias.re.kr, seokkimseok@gmail.com}

\end{center}

\vspace{1cm}

\begin{abstract}

We study supersymmetric AdS$_6$ black holes at large angular momenta, from
the index of 5d SCFTs on $S^4\times\mathbb{R}$ in the large $N$ and Cardy limit.
Our examples are the strong coupling limits of 5d gauge theories on
the D4-D8-O8 system. The large $N$ free energy scales like
$N^{5/2}$, statistically accounting for the entropy of
large black holes in AdS$_6$. Instanton solitons play subtle roles
to realize these deconfined degrees of freedom.

\end{abstract}

\end{titlepage}

\renewcommand{\thefootnote}{\arabic{footnote}}

\setcounter{footnote}{0}

\renewcommand{\baselinestretch}{1}

\tableofcontents

\renewcommand{\baselinestretch}{1.2}

\section{Introduction}

Superconformal field theories (SCFTs) in spacetime dimensions $d>4$ were discovered indirectly
from string theory. First examples are \cite{Witten:1995zh} in 6d, and \cite{Seiberg:1996bd}
in 5d. These QFTs defy microscopic descriptions from traditional Lagrangian
methods so far. One interesting aspect is that they
have much larger numbers of degrees of freedom than conventional gauge theories, at
given gauge group `rank.' For instance, if the QFTs are engineered by
$N$ ($\gg 1$) branes, the 6d/5d QFTs of \cite{Witten:1995zh} and \cite{Seiberg:1996bd,Intriligator:1997pq} exhibit
$N^3$ and $N^{5/2}$ degrees of freedom respectively. This is much larger than $N^2$
for gauge theories on D-branes at weak coupling.

Recently, formulae for certain supersymmetric partition functions for these SCFTs have
been suggested and explored. We shall be interested in the index of 5d SCFTs on
$S^4\times\mathbb{R}$  \cite{Bhattacharya:2008zy}. Its matrix integral formula has been
obtained in \cite{Kim:2012gu}. This formula has been providing new channels to
quantitatively study 5d SCFTs.
In this paper, we shall add one more intriguing finding, by exploring novel
deconfinements of large $N$ 5d SCFTs and the holographically dual black holes in AdS$_6$
spacetime. We study 5d SCFTs engineered on
D4-D8-O8 systems \cite{Seiberg:1996bd,Intriligator:1997pq}.
In this setting, the large $N$ deconfined degrees of freedom
would be visible in the high temperature phase. The gravity dual of deconfinement is
the nucleation of black holes after the Hawking-Page phase transition
\cite{Hawking:1982dh,Witten:1998zw}.
Our deconfined index successfully counts the microstates of the
supersymmetric AdS$_6$ black holes of \cite{Chow:2008ip}, in the framework
of \cite{Choi:2018fdc}.

More precisely, we shall investigate large AdS$_6$ black holes
by studying a Cardy limit of the index. We first note that similar
large BPS black holes in AdS$_5\times S^5$ were accounted for recently, by
exploring the Cardy formula for the index of 4d $\mathcal{N}=4$ Yang-Mills theory
\cite{Choi:2018hmj}. This suggests to resolve a puzzle encountered in an earlier study
\cite{Kinney:2005ej} as follows. In \cite{Kinney:2005ej}, large $N$ and large charge limit
of the index was studied, at order $\mathcal{O}(N^0)$ chemical potentials.
They found no
indication for the macroscopic entropy. The suggestion of \cite{Choi:2018hmj} is
that, the degeneracies at each order in the fugacity expansion are macroscopic
at large charges, but come with rather randomly alternating overall $\pm$ signs. (See also \cite{Choi:2018vbz} for more explanations.) In this situation, if one tries
to Legendre transform the free energy at large charges to extract entropy,
one can generally
expect two possibilities. Since Legendre transformation is insensitive
to the precise quantized values of macroscopic charges, the first possibility is that
it may capture an `averaged'
degeneracy after smearing-out nearby degeneracies with oscillating signs. This may
hugely under-estimate the entropies, as found in \cite{Kinney:2005ej}. Secondly,
it may capture macroscopic degeneracy,
with the oscillating overall signs realized by a rapidly changing phase factor
which can hop between $\pm 1$ as charges are changed by their minimal quanta.
To search for the second possibility, one should admit complexified fugacities at the
saddle point of Legendre transformation. \cite{Choi:2018hmj} considered these complex
fugacities in the Cardy limit, generalizing earlier
studies of \cite{DiPietro:2014bca,Ardehali:2015bla}, precisely accounting for
large AdS$_5$ black holes. Further studies on the Cardy limit and large AdS$_5$
black holes were made in \cite{Honda:2019cio,ardehali}.
Beyond the Cardy limit, \cite{Choi:2018vbz} illustrated the existence of deconfinement
phase transition. In addition, the large $N$ limit of the supersymmetric partition function \cite{Cabo-Bizet:2018ehj} and the index \cite{Benini:2018ywd} were studied, which successfully account for the known black hole solutions at general size.
Cardy free energy was also studied for 6d $(2,0)$ theory on
$S^5\times\mathbb{R}$, counting the dual large BPS black holes in AdS$_7\times S^4$
\cite{Choi:2018hmj}.

We study a similar Cardy limit of the 5d SCFT index in this paper.
We shall take large $N$ limit, and also take the chemical potentials $\omega_{1,2}$
conjugate to the two rotations on $S^4$ to be small. The last limit partly defines our
Cardy limit. See section 2 for the precise definition.
Apparently, the large $N$ calculation in the Cardy limit turns out to be rather simple.
The index on
$S^4\times S^1$ is given by an integral over the $U(1)^r\subset G$ holonomies $\alpha_a$
on $S^1$ which are all periodic variables on a circle,
where $G$ is the 5d gauge group of rank $r$ \cite{Kim:2012gu}.
The integrand consists of a pair of
instanton partition functions \cite{Nekrasov:2002qd}, or more abstractly
the Omega-deformed partition functions of 5d SCFTs in the Coulomb phase.
We seek for the large $N$
saddle points of $\alpha_a$, also taking the Cardy limit $|\omega_{1,2}|\ll 1$.
To get the relevant saddle point, it turns out that one has to complexify $\alpha_a$ to
variables living on cylinders, and spread them over a large interval of length
$\sim N^{\frac{1}{2}}$. This is similar to the partition functions of
5d SCFTs on $S^5$ \cite{Jafferis:2012iv}, and especially to the topologically twisted
indices on suitable spatial manifolds \cite{Hosseini:2018uzp,Crichigno:2018adf,Hosseini:2018usu}
which counted certain black holes in AdS$_6$ \cite{Suh:2018tul,Suh:2018szn}. It seems that the
physical implications of
such novel large $N$ saddle points were not fully discussed in the literature. We find
this especially novel, having in mind the deconfinement phase transition to
$N^{\frac{5}{2}}$ degrees of freedom. The novelty partly has to do with the rather mysterious instanton solitons in higher dimensional gauge theories, concerning their
noncompact internal zero modes and infinite towers of bound states.
We shall comment on these aspects briefly.

The rest of this paper is organized as follows. In section 2,
we study the large $N$ Cardy limit of the index for 5d gauge
theories having AdS$_6$ gravity duals in massive IIA string theory.
These Cardy free energies precisely account for the large
BPS black holes in AdS$_6$, using the recently discovered entropy functions
\cite{Choi:2018fdc} for these black holes. We also  comment on subtle
aspects of our free energy, especially concerning
the physics of instantons. In section 3, we conclude and discuss
some open questions.

\section{Cardy limit of large $N$ 5d SCFTs and black holes}

We first briefly review the large $N$ 5d SCFT models that we shall discuss
in sections 2.1 and 2.2. 5d $\mathcal{N}=1$ SCFTs of our interest live on $N$ D4-branes
probing an O8-plane and $N_f \leq 7$ D8-branes on $\mathbb{C}^2/\mathbb{Z}_n$
\cite{Intriligator:1997pq, Seiberg:1996bd, Bergman:2012kr, Bergman:2013koa}. Note that
$\mathbb{Z}_n$ orbifold is transverse to D4 branes, while O8, D8's are parallel to them.
So in the probe limit, the net spacetime is
given by $\mathbb{R}^{4,1}\times\mathbb{R}^+\times\mathbb{C}^2/\mathbb{Z}_n$,
where $\mathbb{R}^+=\mathbb{R}/\mathbb{Z}_2$ is the direction transverse to the
O8-plane.

\begin{table}[t]
\centering
\hspace*{-0.7cm}
\begin{tabular}{  l || l | l |l }
\hline
orbifold & gauge group & matter & flavor symmetry \\ \hline
$\mathbb{Z}_{2k}^{-}$ & $Sp(N)\!\times \!SU(2N)^{k-1} \!\times\! Sp(N)$ &
$\sum_{i=1}^k({\bf 2N}_i, \overline{\bf 2N}_{i+1})$
& $U(1)_M \!\times \!U(1)_{B}^{k-1} \!\times\! U(1)_I^{k+1}$ \\[0.1cm]
$\mathbb{Z}_{2k}^{+}$ & $SU(2N)^{k}$ &
$\overline{\bf A}_1\!+\!\sum_{i=1}^{k-1}({\bf 2N}_i, \overline{\bf 2N}_{i+1})
\!+\!{\bf A}_k$
& $U(1)_M \!\times\! U(1)_{B}^{k} \!\times\! U(1)_I^{k}$ \\[0.2cm]
$\mathbb{Z}_{2k+1}$ & $Sp(N) \times SU(2N)^{k}$ &
$\sum_{i=1}^k({\bf 2N}_i, \overline{\bf 2N}_{i+1})+{\bf A}_{k+1}$
& $U(1)_M \!\times\! U(1)_{B}^{k} \!\times\! U(1)_I^{k+1}$ \\[0.15cm]
\hline
\end{tabular}
\caption{Properties of the quiver gauge theories. ${\bf A}_i$ denotes rank $2$
antisymmetric hypermultiplet of the $i$'th node, and fundamental matters are not
shown.}\label{quiver}
\end{table}

When $n=1$, the low energy dynamics on D4-branes is described by an
$Sp(N)$ gauge theory with one rank 2
antisymmetric hypermultiplet and $N_f \leq 7$ fundamental hypermultiplets
\cite{Intriligator:1997pq, Seiberg:1996bd}. When $n=2$, there is a $\mathbb{Z}_2$ orbifold.
If it is the orbifold without vector structure, the worldvolume theory on D4-branes is
$SU(2N)$ gauge theory with two rank 2 antisymmetric hypermultiplets and $N_f \leq 7$
fundamental hypermultiplets \cite{Bergman:2012kr, Bergman:2013koa}. The other orbifold
theories for $n\geq 2$ are quiver gauge theories \cite{Bergman:2012kr, Bergman:2013koa}.
Gauge groups, matter contents, and flavor symmetries of these quivers are shown
in Table \ref{quiver}. In all these models,
the $q$'th gauge node of the quiver may have $N_f^{(q)}$ fundamental matters, which
should satisfy $\sum_q N_f^{(q)}=N_f$.
In the table, $\mathbb{Z}_{2k}^{\pm}$ denotes the orbifold without/with vector structure,
respectively. They are associated with two choices for the orientifold projection in
$k$-th twisted sector. $\mathbb{Z}_{2k}^+$ projects onto even states, i.e. this is the `ordinary' orbifold \cite{Bergman:2012kr}.

Bifundamental and antisymmetric fields in these quiver gauge theories can form
various gauge invariant operators: a meson, $n_I$ di-baryons of the bi-fundamental
fields, and $n_A$ Pfaffian baryons of antisymmetric fields. Numbers of the baryon
operators $(n_I, n_A)$ in each quiver gauge theories are given by
$(k,0), \, (k-1,2), \, (k,1)$, respectively. These baryon operators are not all
independent since a product of them is related to the meson operator. The mesonic
$U(1)_M$ symmetry rotates all the antisymmetric and bifundamental fields of the quiver,
with charge $1$ and $2$, respectively.
We shall introduce a mesonic charge $Q_M$, $n_A+n_I$ baryonic charges $Q_{B_{A}}, Q_{B_I}$ and their conjugate chemical potentials
$m, b_{A},b_I$. Then we impose a constraint
\begin{equation}\label{bar-const}
\sum_{A=1}^{n_A} b_A + 2 \sum_{I=1}^{n_I} b_I =0\ ,
\end{equation}
which reduces the number of independent baryonic charges by one. See \cite{Bergman:2012kr}
for more details.

The strong coupling limits of these gauge theories are 5d SCFTs.
In the large $N$ limit, these SCFTs are dual to the massive IIA string theory
in the warped AdS$_6 \times (S^4/\mathbb{Z}_2)/\mathbb{Z}_n$ product background
\cite{Brandhuber:1999np, Bergman:2012kr, Bergman:2013koa}.
The $SU(2)_R$ R-symmetry of the SCFT corresponds to the $SU(2)$ part of
$SU(2) \times U(1)$ isometry of $(S^4/\mathbb{Z}_2)/\mathbb{Z}_n$.
The overall $U(1)_M$ mesonic symmetry, acting on all the antisymmetric and
bifundamental matters, corresponds to the remaining $U(1)$ part of $SU(2) \times U(1)$
isometry. When $n=1, \, 2$, the $U(1)_M$ mesonic symmetry is enhanced to $SU(2)_M$.
This corresponds to the fact that
the isometry of $S^4/\mathbb{Z}_2$ or $(S^4/\mathbb{Z}_2)/\mathbb{Z}_2$
is $SU(2) \times SU(2)$.

The gravity duals of other global symmetries --
$U(1)_B^{n_I+n_A-1}$ baryonic symmetries, $U(1)_I$ instanton symmetries for every
gauge nodes, and flavor symmetries acting on the fundamental matters -- are also well
understood \cite{Intriligator:1997pq, Seiberg:1996bd, Bergman:2012kr, Bergman:2013koa}.
In particular, when $n=1$, $U(1)_I\times SO(2N_f)$ is enhanced to $E_{N_f+1}$
\cite{Seiberg:1996bd, Morrison:1996xf, Douglas:1996xp, Ganor:1996pc, Intriligator:1997pq, Danielsson:1997kt, Hwang:2014uwa} . In the dual gravity, the states charged under $E_{N_f+1}$ are localized at D8-O8, the boundary of $S^4/\mathbb{Z}_2$ 
\cite{Brandhuber:1999np, Bergman:2012kr}. The $SO(2N_f)$ part comes from perturbative
open strings on O8-D8. The $U(1)_I$ charge at $n=1$ is carried by D0-branes
in the gravity dual \cite{Brandhuber:1999np, Bergman:2012kr}. Since the inverse-dilaton
field diverges at the boundary of $S^4/\mathbb{Z}_2$ (i.e. at the 8-brane),
the D0-branes are attracted to the 8-branes and nonperturbatively render
$E_{N_f+1}$ enhancement.

When $n \geq 2$, there can be more $U(1)_I$ instanton symmetries if there are more than
one gauge nodes, and there are $U(1)_B$ baryonic symmetries as well
\cite{Bergman:2012kr, Bergman:2013koa}. The bulk duals of these symmetries are
given as follows \cite{Bergman:2012kr}.
These symmetries basically come from the $\mathbb{Z}_n$ orbifold. $\mathbb{Z}_n$
acts freely on the $S^3$ base of $S^4/\mathbb{Z}_2$, yielding the Lens space $S^3/\mathbb{Z}_n$. The full compact internal space $(S^4/\mathbb{Z}_2)/\mathbb{Z}_n$
has an $A_{n-1}$ singularity at the pole.
There are $n-1$ vanishing 2-cycles at the pole, and also $n-1$ dual finite 2-cycles since the internal space is compact. These cycles should be
identified pairwise by the O8-orientifold.
When $n$ is odd, the O8 action leaves $\frac{n-1}{2}$ vanishing 2-cycles and $\frac{n-1}{2}$ finite 2-cycles. When $n$ is even, we should be careful about the $\frac{n}{2}$-th
unpaired 2-cycles. If the $\mathbb{Z}_{n}$ orbifold is without vector structure, $\frac{n}{2}$-th vanishing 2-cycle is projected out, while $\frac{n}{2}$-th finite 2-cycle is mapped to itself. So there are $\frac{n-2}{2}$ vanishing 2-cycles and $\frac{n}{2}$ finite 2-cycles after the O8 projection. On the contrary, when $\mathbb{Z}_{n}$ is the orbifold with vector structure, $\frac{n}{2}$-th vanishing 2-cycle is mapped to itself, while $\frac{n}{2}$-th finite cycle is projected out, leaving $\frac{n}{2}$ vanishing 2-cycles and $\frac{n-2}{2}$ finite 2-cycles after O8-projection.
Baryons are described by D2-branes wrapping the finite 2-cycles.
Instantons are dual to D0-brane and D2-branes wrapping the vanishing 2-cycles, i.e. fractional D0-branes. These explain
all the symmetries listed in Table \ref{quiver}.

\subsection{$Sp(N)$ theories}\label{sec: Cardy-Sp(N)}

In this subsection, we study the large $N$ index for the 5d $\mathcal{N}=1$
gauge theories with $Sp(N)$ gauge group, one rank 2 antisymmetric hypermultiplet, and
$N_f \leq 7$ fundamental hypermultiplets \cite{Kim:2012gu}.
We shall consider the radially quantized theory on $S^4 \times \mathbb{R}$. We choose
a supercharge $Q$ to define the index, so that we count
$\frac{1}{8}$-BPS states annihilated by the supercharge $Q$ and its conjugate conformal
supercharge $S=Q^\dagger$. We will denote by $j_1,\, j_2$ the Cartan generators of
$SU(2)_1 \times SU(2)_2 \subset Sp(2) \cong SO(5)$ rotation symmetry, and by $R$
the Cartan generator of $SU(2)_R$ R-symmetry. We introduce the fugacities
$e^{-\beta},\, x,\, y$
for $\{Q,S\},\, j_1+R,\, j_2$ in $F(4)$ superconformal symmetry, which commute with the supercharges $Q$ and $S$. Since the antisymmetric representation of $Sp(N)$ group is real, the antisymmetric hypermultiplet splits into two half-hypermultiplets, which transform as a doublet under
$Sp(1)_M \cong SU(2)_M$ global symmetry. We call its Cartan generator $h$.
This system also has $SO(2N_f)$ flavor symmetry rotating the fundamental quarks.
We call their Cartan generators $H_l$, with $l=1,\cdots,N_f$.
Finally, there is a $U(1)_I$ topological symmetry related to the current
$j_\mu\sim\star_5{\rm tr}(F\wedge F)_\mu$. The corresponding conserved
charge is the instanton number $k$.
We introduce the fugacities $e^{- m}, \, e^{-M_l}$'s, and $q$ for
$h$, $H_l$, $k$. The index is defined as \cite{Bhattacharya:2008zy,Kim:2012gu}
\begin{equation}\label{SpN-index-trace}
Z(x,y,m,M_l,q)=\textrm{Tr}\left[(-1)^F e^{-\beta\{Q,S\}} x^{2(j_1+R)} y^{2j_2} e^{-m h} e^{-\sum_l M_l H_l} q^k \right]\ ,
\end{equation}
where $F$ is the fermion number operator. The trace is taken over the Hilbert space
of the QFT on $S^4\times\mathbb{R}$. This index counts BPS states, for which the eigenvalue
of $\{Q,S\}$ is $0$. So the index does not depend on $\beta$.
For the $Sp(N)$ theory with one rank 2 antisymmetric hypermultiplet and $N_f$ fundamental hypermultiplets, this index is given by \cite{Kim:2012gu}
\begin{eqnarray}\label{Sp(N)-index}
Z(x,y,m,M_l,q)&=&\oint \left[d\alpha\right] \textrm{PE}\left[f_{vec}(x,y,e^{i\alpha_a})+f_{mat}^{asym}(x,y,e^{i\alpha_a},
e^{m})+f_{mat}^{fund}(x,y,e^{i\alpha_a},e^{M_l})\right]\nonumber\\
&& \qquad \quad \times \prod_{\pm}Z_{\rm inst}
(x^{\pm 1},y^{\pm 1},e^{\pm i\alpha_a},e^{\pm m},e^{\pm M_l},q^{\pm 1})\ ,
\end{eqnarray}
where $[d\alpha]$ including the Haar measure is given by
\begin{equation}
\left[d\alpha \right]=\frac{1}{2^NN!}\left[\prod_{a=1}^N \frac{d\alpha_a}{2\pi}
\left(2\sin \alpha_a\right)^2 \right] \prod_{a<b} \left[2 \sin \left( \frac{\alpha_a-\alpha_b}{2} \right)\right]^2 \left[2 \sin \left( \frac{\alpha_a+\alpha_b}{2} \right) \right]^2\ .
\end{equation}
$f_{vec},\, f_{mat}^{asym},\, f_{mat}^{fund}$ are the letter indices for the vector, antisymmetric, and fundamental hypermultiplets, given by
\begin{eqnarray}\label{letter-index}
f_{vec}&=& - \frac{x(y+1/y)}{(1-xy)(1-x/y)} \left[\sum_{a<b}^Ne^{i(\pm \alpha_a\pm\alpha_b)}
+ \sum_{a=1}^N \left(e^{-2i\alpha_a} + e^{2i\alpha_a} \right) + N \right] \ , \\
f_{mat}^{asym}&=&
\frac{x}{(1-xy)(1-x/y)} (e^{m/2}+e^{-m/2}) \left[\sum_{a<b}^N
e^{i(\pm\alpha_a\pm\alpha_b)} +N \right] \ , \nonumber\\
f_{mat}^{fund}&=& \frac{x}{(1-xy)(1-x/y)} \sum_{a=1}^N \sum_{l=1}^{N_f} \left(e^{-i\alpha_a-M_l}+e^{i\alpha_a-M_l}+e^{-i\alpha_a+M_l}
+e^{i\alpha_a+M_l} \right)\ .\nonumber
\end{eqnarray}
Our notation is that the terms with $\pm$ are all summed over: for instance,
$e^{i(\pm\alpha_a\pm\alpha_b)}\equiv e^{-i\alpha_a-i\alpha_b}+e^{-i\alpha_a+i\alpha_b}+e^{i\alpha_a-i\alpha_b}
+e^{i\alpha_a+i\alpha_b}$. $Z_{\rm inst}$ is the Coulomb branch
instanton partition function \cite{Nekrasov:2002qd}, taking the form of
\begin{equation}
Z_{\rm inst}=\sum_{k=0}^\infty q^k Z_k(x, \, y, \, e^{i\alpha_a}, \,
e^{m}, \, e^{M_l})\ , \quad Z_{k=0}=1\ ,
\end{equation}
where $Z_k$ is the $k$-instanton contribution. $Z_k$ can be computed using
the methods of \cite{Nekrasov:2004vw, Shadchin:2005mx,Hwang:2014uwa}.
PE in (\ref{Sp(N)-index}) is the Plethystic exponential defined as
\begin{equation}
\textrm{PE}\left[f(\{t\})\right] = \exp \left[\sum_{n=1}^\infty \frac{1}{n} f(\{t^n\}) \right]\ ,
\end{equation}
where $\{t\}$ collectively denotes all fugacities for gauge and global symmetries appearing in $f$.

For later convenience, we redefine the fugacities as $e^{-\omega_1}, \, e^{-\omega_2}$ for
the angular momenta $J_1\equiv j_1+j_2, \, J_2\equiv j_1-j_2$, which act on the orthogonal 2-planes of $\mathbb{R}^5$ which contains $S^4$. We also define
$e^{-\tilde{\Delta}}\equiv e^{-(\Delta-2\pi i) }$ for $R$.
They are related to the original fugacities as
\begin{equation}
\begin{aligned}
e^{-\omega_1}=xy, \; e^{-\omega_2}=x/y, \; e^{-\tilde{\Delta}} = x^2\ .
\end{aligned}
\end{equation}
The new chemical potentials satisfy $\Delta - \omega_1 -\omega_2 = 2\pi i$
(mod $4\pi i$).
Since $R,J_1,J_2$ are normalized to be half-integers, all chemical potentials
have $4\pi i$ periods. This is the reason for the mod $4\pi i$ in the last equation.
Below, we shall always take
\begin{equation}\label{index-condition}
  \Delta - \omega_1 -\omega_2 = 2\pi i\ .
\end{equation}
The Haar measure can be rewritten as
\begin{equation}
\left[d\alpha \right] = \frac{1}{2^N N!} \prod_{a=1}^N \frac{d\alpha_a}{2\pi} \; \textrm{PE} \left[-\sum_{a<b}^N e^{i(\pm\alpha_a\pm\alpha_b)}
- \sum_{a=1}^N \left(e^{-2i\alpha_a} + e^{2i\alpha_a} \right) \right]\ .
\end{equation}
Combining this exponent of PE from the Haar measure to $f_{\rm vec}$, one obtains
\begin{equation}\label{Sp(N)-vec}
\begin{aligned}
\tilde{f}_{vec} = & - \frac{1+x^2}{(1-xy)(1-x/y)} \left[\sum_{a<b}^N e^{i(\pm\alpha_a\pm\alpha_b)} + \sum_{a=1}^N \left(e^{-2i\alpha_a} + e^{2i\alpha_a} \right) \right]
+ \left(1 - \frac{1+x^2}{(1-xy)(1-x/y)} \right) N \\
= & - \frac{1+e^{-\tilde{\Delta}}}{(1-e^{-\omega_1})(1-e^{-\omega_2})} \left[\sum_{a<b}^N
e^{i(\pm\alpha_a\pm\alpha_b)} + \sum_{a=1}^N \left(e^{-2i\alpha_a} + e^{2i\alpha_a} \right)
\right] + \left(1 - \frac{1+e^{-\tilde{\Delta}}}{(1-e^{-\omega_1})(1-e^{-\omega_2})} \right) N \\
=& - \frac{2\cosh \frac{\tilde{\Delta}}{2}}{2\sinh \frac{\omega_1}{2} \cdot 2\sinh \frac{\omega_2}{2}} \left[\sum_{a<b}^N e^{i(\pm\alpha_a\pm\alpha_b)}
+ \sum_{a=1}^N \left(e^{-2i\alpha_a} + e^{2i\alpha_a} \right) \right]
+\left(1 - \frac{2\cosh \frac{\tilde{\Delta}}{2}}{2\sinh \frac{\omega_1}{2} \cdot 2\sinh \frac{\omega_2}{2}} \right) N\ .
\end{aligned}
\end{equation}
We used \eqref{index-condition} on the last line. Other letter
indices are given by
\begin{equation}\label{Sp(N)-mat}
  f_{mat}^{asym}= \frac{2 \cosh \frac{m}{2}}{2\sinh \frac{\omega_1}{2} \cdot 2\sinh \frac{\omega_2}{2}} \left[\sum_{a<b}^N e^{i(\pm\alpha_a\pm\alpha_b)}+N \right] \ \ ,\ \
  f_{mat}^{fund}= \frac{\sum_{l=1}^{N_f} 2 \cosh M_l}{2\sinh \frac{\omega_1}{2} \cdot 2\sinh \frac{\omega_2}{2}} \sum_{a=1}^N \left(e^{i\alpha_a}+e^{-i\alpha_a} \right)\ .
\end{equation}

Now we consider a Cardy-like limit $|\omega_i| \ll 1$
\cite{Choi:2018hmj}. We will keep $\omega_i$'s complex with
$\textrm{Re} (\omega_i)>0$. Due to
\eqref{index-condition}, $\Delta$ will be close to $2\pi i$.
Namely, its imaginary part is $\mathcal{O}(1)$ and close to $2\pi i$, while its
real part is small at order $|\omega_i|$.
However, as in \cite{Choi:2018hmj}, for convenient intermediate manipulations, we shall temporarily take $\Delta$ to be pure imaginary, and continue back to a complex number with
small real part later. The other chemical potentials $m, \, M_l$'s are kept purely imaginary
(which may be continued back to suitable complex numbers later, if one wishes).
Then following the similar procedures used in \cite{Choi:2018hmj}, ignoring
the Cartan parts of $Sp(N)$ at large $N$, the PE of the letter indices are approximated as
{\allowdisplaybreaks
\begin{align}
&\textrm{PE} \left[\tilde{f}_{vec} \right] \sim \exp \Bigg[ - \frac{1}{\omega_1 \omega_2} \sum_{n=1}^\infty \frac{e^{\frac{n(\Delta-2\pi i)}{2}}+e^{-\frac{n(\Delta-2\pi i)}{2}}}{n^3} \Bigg( \sum_{a=1}^N \left(e^{-2in\alpha_a} + e^{2in\alpha_a} \right)
+ \sum_{a<b}^N
e^{i(\pm\alpha_a\pm\alpha_b)}\Bigg) \Bigg] \nonumber\\
&= \exp \Bigg[ - \frac{1}{\omega_1 \omega_2} \sum_{n=1}^\infty \frac{(- e^{\frac{\Delta}{2}})^n+(-e^{-\frac{\Delta}{2}})^n}{n^3} \Bigg( \sum_{a=1}^N \left(e^{-2in\alpha_a} + e^{2in\alpha_a} \right)
+ \sum_{a<b}^N e^{i(\pm\alpha_a\pm\alpha_b)}\Bigg) \Bigg] \nonumber\\
&= \exp \Bigg[ - \frac{1}{\omega_1 \omega_2} \Bigg( \sum_{a=1}^N
\textrm{Li}_3(-e^{\pm \frac{\Delta}{2}\pm 2i\alpha_a})
+ \sum_{a<b}^N  \textrm{Li}_3(-e^{\pm \frac{\Delta}{2}\pm i\alpha_a\pm i\alpha_b})
\Bigg) \Bigg]
\equiv \exp \left[ - \frac{\mathcal{F}_{vec}(\alpha_a, \, \Delta)}{\omega_1 \omega_2}\right]
\ , \\
&\textrm{PE} \left[f_{mat}^{asym} \right] \sim \exp \Bigg[ \frac{1}{\omega_1 \omega_2} \sum_{n=1}^\infty \frac{1}{n^3} (e^{\frac{nm}{2}}+e^{-\frac{nm}{2}}) \sum_{a<b}^N
e^{i(\pm\alpha_a\pm\alpha_b)} \Bigg] \nonumber\\
&= \exp \Bigg[ \frac{1}{\omega_1 \omega_2} \sum_{a<b}^N
\textrm{Li}_3(e^{\pm\frac{m}{2}\pm i\alpha_a\pm i\alpha_b})\Bigg]
\equiv \exp \left[ - \frac{\mathcal{F}_{mat}^{asym}(\alpha_a, \, m)}{\omega_1 \omega_2}\right]\ , \\
& \textrm{PE} \left[f_{mat}^{fund} \right] \sim \exp \Bigg[ \frac{1}{\omega_1 \omega_2} \sum_{n=1}^\infty \frac{1}{n^3} \sum_{l=1}^{N_f} (e^{nM_l}+e^{-nM_l}) \sum_{a=1}^N \left(e^{in\alpha_a}+e^{-in\alpha_a} \right) \Bigg] \nonumber\\
&= \exp \Bigg[ \frac{1}{\omega_1 \omega_2} \sum_{l=1}^{N_f} \sum_{a=1}^N \sum_{\pm} \left(\textrm{Li}_3(e^{\pm M_l+i\alpha_a})+\textrm{Li}_3(e^{\pm M_l-i\alpha_a}) \right)\!\Bigg]\!
\equiv \exp \left[ - \frac{\mathcal{F}_{mat}^{fund}(\alpha_a, \, M_l)}{\omega_1 \omega_2}
\right] .
\end{align}
}
Here, we used the power series definition of the trilogarithm function \eqref{power-trilog}.
$\Delta$ here can be taken back to be the one satisfying (\ref{index-condition}).
It is important to remember that the imaginary parts of chemical potentials
may be kept nonzero and $\mathcal{O}(1)$, to obstruct boson/fermion cancelations as
in \cite{Choi:2018fdc,Choi:2018hmj}. (Especially, those of $\Delta$ and $m$ will play
important roles later.) On the other hand,
the real parts in the Cardy limit are either kept small (for $\Delta$) or just set to
$0$ (for $m,M_l$, since we are uninterested in such continuations).
The integral contours for the variables $e^{i\alpha_a}$'s are all along the unit circle,
$|e^{i\alpha_a}|=1$.

The instanton part $Z_{\rm inst}$ is subtler, and needs a more careful study.
So far, $Z_{\rm inst}$ is understood only as a series expansion in certain fugacity.
Canonically, the fugacity $q$ for the $U(1)_I$ flavor symmetry is the expansion
parameter of $Z_{\rm inst}$. As we shall consider
dual AdS$_6$ black holes which do not carry flavor charges, we set $q=1$ (or to
a generic order $1$ value so that it does not provide an expansion parameter).
This qualitatively corresponds to taking the 5d gauge coupling to infinity.
So apparently, the series which sums over the instanton number $k$ is unsuppressed.
The proper way of understanding (\ref{SpN-index-trace}), (\ref{Sp(N)-index}) 
was explained in \cite{Kim:2012gu},
as a series expansion in the fugacity $x$. However, in our Cardy limit, we take
$|x|\rightarrow 1^{-}$, so that it is unclear how to understand
$Z_{\rm inst}$ part.
Here, we quote an idea explored in \cite{Jafferis:2012iv}, which is to focus on a
particular large $N$ saddle point of $N$ integral variables. The integral variables
of \cite{Jafferis:2012iv} are $N$ real scalars $\phi$, while
the analogous $N$ variables in our problem will be analytically continued $\alpha_a$'s
in their imaginary directions. (Namely, $-i\alpha_a>0$ with purely imaginary $\alpha_a$'s
will play the role of $\phi$ of \cite{Jafferis:2012iv}.)
\cite{Jafferis:2012iv} considered a saddle point in which the $N$ scalars are spread
with a wide width $N^{\frac{1}{2}}$ (which is assumed to be the dominant one), and
self-consistently showed that the instanton parts can be approximated to
$Z_{\rm inst}\approx 1$. A simple argument for ignoring the instanton part was presented
in \cite{Jafferis:2012iv}, based on the renormalized gauge coupling in the Coulomb branch
of 5d SCFT. In the next paragraph and in appendix B, we shall correct some naive
1-loop arguments in \cite{Jafferis:2012iv} made for this conclusion.
However, this will not spoil their final conclusion that $Z_{\rm inst}\approx 1$.

The idea of \cite{Jafferis:2012iv} is that, if the scalar VEV $\phi$ is nonzero,
there is a nonzero 1-loop contribution to the 5d gauge coupling in the Coulomb branch.
The 1-loop effective coupling which depends on the scalar schematically takes the form of
\begin{equation}\label{effective-coupling}
  \frac{1}{g_{\rm eff}^2(\phi)}\sim(8-N_f)|\phi|
\end{equation}
at infinite bare coupling (corresponding to $q=1$ in our setting). This expression
is in fact slightly imprecise. This is because $\frac{1}{g_{\rm eff}^2(\phi)}$ is the
coefficient of the kinetic term of the Coulomb branch fields, so should be
an $N\times N$ matrix in our $Sp(N)$ theory. The above expression should be understood
as a schematic expression for the eigenvalues, where $\phi$ denotes a component of the
$N$ Coulomb VEVs. The key argument of \cite{Jafferis:2012iv} is that the mass of instantons
is basically the inverse-square of the gauge coupling, so in the Coulomb branch it should
also acquire a contribution of the form (\ref{effective-coupling}).
If this is the case, and if the saddle point values of $\phi$'s are large,
the $k$ instanton correction would come with a suppression factor of
$\sim e^{-k(8-N_f)|\phi|}$, with $|\phi|\sim
N^{\frac{1}{2}}$. This was the argument for self-consistently approximating
$Z_{\rm inst}\approx 1$. However, we find that such a 1-loop argument is incomplete,
for the following reason. In the brane setting, $\frac{1}{g_{\rm eff}^2(\phi)}$ is
given by the running dilaton field sourced by O8-D8, where $\phi$ is the
coordinate for the transverse direction to the 8-branes. If a D4-brane is at the
location $\phi$, the D0-brane ($\sim$ instanton) bound to it will find
$\frac{1}{g_{\rm eff}^2(\phi)}$ is its mass. However, if one studies the detailed
structures of $Z_{\rm inst}$ for this system, D0-branes can be stuck not only to
$N$ separated D4-branes. They can also be bound to the O8-plane at $\phi=0$, still
contributing to the 5d QFT spectrum. Details will be explained in appendix B.
Therefore, the 1-loop argument of regarding
$\frac{1}{g_{\rm eff}^2(\phi)}$ as the instanton mass and the suppression factor
is incorrect. Had such a claim been true, one would have expected the suppression factor
of $e^{-k(8-N_f)|\phi|}$ at $k$ instanton sector, with real
$\phi\equiv -i\alpha\sim N^{\frac{1}{2}}$. However, as explained in appendix B,
the D0-branes bound to O8 turns out to have lighter quantum masses, so that
the true suppression factor for $k$ instantons turns out to be
\begin{eqnarray}
  N_f\neq 0&:&\sim e^{-k|\phi|}\\
  N_f=0&:&\sim e^{-2k|\phi|}\ .\nonumber
\end{eqnarray}
For most values of $N_f$, this is larger than the naively estimated suppression
factors.\footnote{More fundamentally, such exotic masses are allowed since
the instanton masses cannot be determined just from the 5d effective action in
the Coulomb branch. For instance, the argument above for D0-D4 bounds uses
string theory. SK thanks Hee-Cheol Kim, Kimyeong Lee and Gabi Zafrir for related
discussions in 2017.}
In any case, although the detailed estimates in the literature are
incorrect, the final conclusion $Z_{\rm inst}\approx 1$ will not change.
Among the $N$ eigenvalues $\alpha_a$, most of them will take large imaginary
values $\propto N^{\frac{1}{2}}$, so that the above suppression factors are indeed
small. $\alpha_a$ will not be large for some eigenvalues,
but their number is much smaller than $N$ so that the leading
large $N$ free energy will not be affected \cite{Jafferis:2012iv}.

In our problem, we shall consider the large $N$ and Cardy limit $\omega_{1,2}\rightarrow 0$
together, seeking for a similar saddle point. Our large $N$ saddle point will complexify
the angle variables $\alpha_a$, into cylinders. The complexified
$\alpha_a$'s will be spread at order $\mathcal{O} (\sqrt{N})$ in their imaginary directions.
This is very similar to the studies made with the 5d topologically twisted indices \cite{Hosseini:2018uzp,Crichigno:2018adf}.
Therefore, with such spreading of eigenvalues assumed (to be shown later in this section),
the instanton contribution to the free energy is exponentially suppressed
at large $N$.
So we shall ignore the instanton contribution to $\log Z$ from now.
More comments on these large $N$ saddle point, and the subtle roles of
$Z_{\rm inst}$, will be postponed to  section 2.4.
With these understood, approximately setting $Z_{\rm inst}\approx 1$,
one obtains the following expression for the large $N$ Cardy index at
$|\omega_i| \ll 1$:
\begin{equation}\label{Cardy-index}
\begin{aligned}
Z(\omega_1, \, \omega_2, \, \Delta, \, m, \, M_l) &\sim \frac{1}{2^N N!} \oint \prod_{a=1}^N \frac{d\alpha_a}{2\pi} \exp \left[ - \frac{\mathcal{F}^{pert}(\alpha_a,\, \Delta, \, m, \, M_l)}{\omega_1 \omega_2}  \right] \\
&\equiv \frac{1}{2^N N!} \oint \prod_{a=1}^N \frac{d\alpha_a}{2\pi} \exp \left[ - \frac{\mathcal{F}_{vec}+\mathcal{F}_{mat}^{asym}+\mathcal{F}_{mat}^{fund}}{\omega_1 \omega_2}  \right]\ ,
\end{aligned}
\end{equation}
where $\mathcal{F}^{pert}$ is the perturbative part of the effective action.

At $|\omega_{1,2}| \to 0$ and $N\gg 1$, one can evaluate \eqref{Cardy-index}
by a saddle point method. We assume that the eigenvalues are spread
at order $\mathcal{O} (N^\alpha)$, with $0<\alpha<1$, in the imaginary direction
at large $N$ \cite{Jafferis:2012iv, Hosseini:2018uzp,Crichigno:2018adf}. The ansatz for the
eigenvalue distribution is given by
\begin{equation}\label{ansatz}
\alpha_a= i N^\alpha x_a \quad (0<x_a<x_*)\ .
\end{equation}
Here, $x_a$'s are of order $\mathcal{O}(N^0)$, and the value of $\alpha$
will be determined later. We restricted $\textrm{Im} (\alpha_a)>0$ and
 also ordered $x_a$'s to be increasing, using the Weyl symmetry
of $Sp(N)$, setting $0<x_1<x_2<\cdots < x_N$.
Since we assume $0<\alpha<1$, the $N$ eigenvalues will be densely distributed on an interval of length $\sim N^\alpha$. The range $(0,x_*)$ will be determined later. We take the continuum limit by defining the continuous variable $x\in(0,x_\ast)$
and introducing the density function of eigenvalues $\rho(x)=\frac{1}{N} \frac{da}{dx}$
normalized as $\int \rho(x) dx =1$. Then we replace the sum over $a$ by an integral
of the form
\begin{equation}\label{continuum}
\sum_{a=1}^N \; \to \; N \int_0^{x_*} dx \rho(x)\ ,
\end{equation}
in the $N \to \infty$ continuum limit.

Before proceeding, we note again that the chemical potentials $m,M_l$ all have $4\pi i$
periodicity. We shall assume that all parameters $m,M_l$ are purely imaginary,
and put them in the `canonical chamber' $(0, 4\pi i)$. The expressions of the
free energy in different chambers can be found by periodic shifts of the variables.
Applying the ansatz \eqref{ansatz} and taking the continuum limit \eqref{continuum}, the
contribution of the antisymmetric hypermultiplet
is given by
\begin{eqnarray}
\mathcal{F}_{mat}^{asym} &=& - N^2 \int_0^{x_*} dx \rho(x) \int_x^{x_*} dx' \rho(x') \sum_\pm \Big[\textrm{Li}_3 (e^{-N^\alpha (x+x') \pm \frac{m}{2}}) + \textrm{Li}_3 (e^{-N^\alpha (-x+x') \pm \frac{m}{2}}) \\
&& \qquad \qquad \qquad \qquad \qquad \qquad \qquad \quad + \textrm{Li}_3 (e^{N^\alpha (x+x') \pm \frac{m}{2}}) + \textrm{Li}_3 (e^{N^\alpha (-x+x') \pm \frac{m}{2}}) \Big]
\nonumber\\
&\sim & \frac{2}{3} N^{2+\alpha} \int_0^{x_*} dx \rho(x) \int_x^{x_*} dx' \rho(x') \left[\left(3\frac{m}{2}\left(\frac{m}{2}-2\pi i\right) - 2\pi^2\right)x' +N^{2\alpha} (x'^3+3x^2x') \right]\ ,\nonumber
\end{eqnarray}
where $m$ is in the range $(0,4\pi i)$, as explained above. Here, we used the trilogarithm formulae \eqref{trilog-zero}, \eqref{trilog-large} at $N \to \infty$. Similarly, the fundamental hypermultiplet contribution is given by
\begin{eqnarray}
  \mathcal{F}_{mat}^{fund} &=& -N \sum_{l=1}^{N_f} \int_{0}^{x_*} dx \rho (x) \sum_{\pm} \left[\textrm{Li}_3 (e^{-N^\alpha x \pm M_l}) + \textrm{Li}_3 (e^{N^\alpha x \pm  M_l})\right] \nonumber\\
&\sim & \frac{1}{3} N^{1+3\alpha} \sum_{l=1}^{N_f} \int_{0}^{x_*} dx \rho (x) x^3
= \frac{N_f}{3} N^{1+3\alpha} \int_{0}^{x_*} dx \rho (x) x^3\ .
\end{eqnarray}
Finally, the vector multiplet contribution is given by
\begin{eqnarray}\label{F-vec}
\mathcal{F}_{vec} &=& N \int_0^{x_*} dx \rho(x) \sum_{\pm} \left[\textrm{Li}_3 (-e^{-2N^\alpha x\pm \frac{\Delta}{2}}) + \textrm{Li}_3 (-e^{2N^\alpha x\pm \frac{\Delta}{2}})\right] \\
&&+ N^2 \int_0^{x_*} dx \rho(x) \int_x^{x_*} dx' \rho(x') \sum_{\pm} \Big[\textrm{Li}_3 (-e^{-N^\alpha (x+x')\pm \frac{\Delta}{2}}) + \textrm{Li}_3 (-e^{-N^\alpha (-x+x')\pm \frac{\Delta}{2}}) \nonumber\\
&&\qquad \qquad \qquad \qquad \qquad \qquad \quad + \textrm{Li}_3 (-e^{N^\alpha (x+x')\pm \frac{\Delta}{2}}) + \textrm{Li}_3 (-e^{N^\alpha (-x+x')\pm \frac{\Delta}{2}}) \Big]
\nonumber\\
&\sim & -\frac{8}{3} N^{1+3\alpha} \int_0^{x_*} dx \rho(x) x^3 \nonumber\\
&&+ N^2 \int_0^{x_*} dx \rho(x) \int_x^{x_*} dx' \rho(x') \sum_{\pm} \Big[\textrm{Li}_3 (e^{N^\alpha (x+x')\pm \frac{\omega_1+\omega_2}{2}}) + \textrm{Li}_3 (e^{N^\alpha (-x+x')\pm \frac{\omega_1+\omega_2}{2}}) \Big] \nonumber\\
&\sim & -\frac{8}{3} N^{1+3\alpha} \int_0^{x_*} dx \rho(x) x^3 \nonumber\\
&&- \frac{2}{3} N^{2+\alpha}\! \int_0^{x_*}\!\!\!dx \rho(x) \int_x^{x_*}
\!\!\!dx' \rho(x') \left[\left(3\left(\frac{\Delta}{2} + \pi i\right)\left(\frac{\Delta}{2} - \pi i\right)
\!-\! 2\pi^2\right)x'  +N^{2\alpha} (x'^3\!+\!3x^2x') \right]\ ,\nonumber
\end{eqnarray}
where $\Delta\approx 2\pi i$, and we also used \eqref{index-condition}.\footnote{Here
we applied \eqref{trilog-large} valid for $-2\pi <-i\Delta<2\pi$. The value
$\Delta\approx 2\pi i$ constrained by (\ref{index-condition}) is actually close to the
edge of this interval, so one might wonder if using this formula is valid.
We performed a similar calculus for $2\pi<-i\Delta<6\pi$ and confirmed the continuity
of $\mathcal{F}_{vec}$ at $\Delta=2\pi i$, so that using (\ref{F-vec}) near $\Delta=2\pi i$
is fine.}
Note that in the final expressions of
$\mathcal{F}_{mat}^{asum}$, $\mathcal{F}_{vec}$, the last terms
$\sim N^{2\alpha}(x^{\prime 3}+3x^2x^\prime)$ in the integrand
look more dominant than the remaining terms. We keep
the apparently subdominant terms in foresight, as they will be dominant after
a partial cancelation at the saddle point.

Collecting all, one obtains
\begin{eqnarray}\label{eff-pre}
\mathcal{F}^{pert} &\sim&
- \frac{8-N_f}{3} N^{1+3\alpha} \int_{0}^{x_*} dx \rho (x) x^3 + 2 \gamma  N^{2+\alpha} \int_0^{x_*} dx \rho(x) \int_x^{x_*} dx' \rho(x') x'\ , \\
\gamma &\equiv& \frac{m}{2}\left(\frac{m}{2}-2\pi i \right) - \left(\frac{\Delta}{2} + \pi i\right)\left(\frac{\Delta}{2} - \pi i\right) >0 \ \ \
\textrm{ with }\ 0 < -im < 4\pi\ , \;\; \Delta \approx 2\pi i\ .
\nonumber
\end{eqnarray}
For later convenience, we will use the following alternative expression for the last integral:
\begin{eqnarray}
\int_0^{x_*} dx \rho(x) \int_x^{x_*} dx' \rho(x') x' &=& \int_0^{x_*} dx'
\int_0^{x'} dx \, \rho(x) \rho(x') x'
= \int_0^{x_*} dx \int_0^{x} dx' \, \rho(x') \rho(x) x \nonumber\\
&=& \frac{1}{2} \left[\int_0^{x_*} dx \int_0^{x} dx' \, \rho(x) \rho(x') x
+ \int_0^{x_*} dx \int_x^{x_*} dx' \, \rho(x) \rho(x') x' \right] \nonumber\\
&=& \frac{1}{2} \int_0^{x_*} \int_0^{x_*} dx \, dx' \rho(x) \rho(x') \, \frac{x+x'+|x-x'|}{2}\ .
\end{eqnarray}

We extremize \eqref{eff-pre} in $\rho(x)$, where $\rho(x)$ is nonzero only in
$0<x<x_\ast$, and satisfies $\int_{0}^{x_*} \rho(x) dx =1$, $\rho(x)\geq 0$.
To find a nontrivial saddle point at $N\rightarrow\infty$, all
terms should be of the same order in $N$. So we set
\begin{equation}
N^{1+3\alpha}=N^{2+\alpha} \; \rightarrow \; \alpha = \frac{1}{2}\ ,
\end{equation}
which implies $\mathcal{F} \propto N^{\frac{5}{2}}$. Also note that this
setting justifies ignoring the instanton corrections, as
explained earlier in this section. Introducing the Lagrange multiplier $\lambda$ for the constraint $\int_0^{x_*} \rho(x) dx =1 $, one should extremize
\begin{equation}\label{eff-pre-2}
\mathcal{F} = N^{\frac{5}{2}} \bigg[ - \lambda \left(\int_{0}^{x_*}\!\!\!\rho(x) dx -1 \right) - \frac{8\!-\!N_f}{3} \int_{0}^{x_*}\!\!dx \rho (x) x^3
+\gamma \int_0^{x_*}\!\!\int_0^{x_*}\!\!dx \, dx' \rho(x) \rho(x') \, \frac{x+x'+|x-x'|}{2} \bigg]\ .
\end{equation}
The other constraint $\rho(x) \geq 0$ is to be confirmed later, after obtaining the extremal solution. Extremizing this functional with $\rho(x)$, one obtains
\begin{equation}\label{saddle-eq}
\lambda + \frac{8-N_f}{3} x^3 = \gamma \int_0^{x_*} dx' \rho(x') \left(x+x'+|x-x'|\right)
= 2\gamma \left[x \int_0^{x} dx' \rho(x') + \int_x^{x_*} dx' \rho(x') x' \right]\ .
\end{equation}
Differentiating this equation with $x$, one obtains
\begin{equation}
(8-N_f) x^2 = 2\gamma \left[\int_0^x dx' \rho(x') + x \rho(x) - \rho(x) x \right]
= 2\gamma \int_0^x dx' \rho(x')\ .
\end{equation}
Differentiating once more with $x$, one obtains
\begin{equation}\label{extremal-den}
\rho(x) = \frac{8-N_f}{\gamma} \, x \geq 0 \quad (x \in [0,x_*])\ .
\end{equation}
$\rho(x)$ is always positive for $x>0$, since $\gamma>0$ and $N_f \leq 7$. From  $\int_0^{x_*} \rho(x) dx = 1$, one obtains
\begin{equation}\label{extremal-bound}
x_* = \sqrt{\frac{2\gamma}{8-N_f}}\ .
\end{equation}
$\lambda$ is determined by \eqref{saddle-eq}, whose solution is
$\lambda = \frac{2(8-N_f)}{3}x_*^3$. Inserting these solutions
\eqref{extremal-den}, \eqref{extremal-bound} into \eqref{eff-pre-2}, one obtains
\begin{equation}
\mathcal{F} \sim \frac{8\sqrt{2}}{15} \frac{N^{\frac{5}{2}}}{\sqrt{8-N_f}} \gamma^{\frac{3}{2}}\ .
\end{equation}
So the large $N$ and Cardy free energy is given by
\begin{equation}\label{asymp-free-energy}
\log Z \sim - \frac{8\sqrt{2}}{15} \frac{N^{\frac{5}{2}}}{\sqrt{8-N_f}} \frac{\gamma^{\frac{3}{2}}}{\omega_1 \omega_2}\ ,
\end{equation}
where
\begin{equation}
\gamma = \frac{m}{2}\left(\frac{m}{2}-2\pi i \right) - \left(\frac{\Delta}{2} + \pi i\right)\left(\frac{\Delta}{2} - \pi i\right)
= - \frac{(\Delta+ \hat{m})(\Delta - \hat{m})}{4}>0\ ,
\end{equation}
where we defined $\hat{m} = m-2\pi i$.

To summarize, the large $N$ and Cardy free energy of the index
\eqref{asymp-free-energy} is given by
\begin{equation}\label{free-energy}
\log Z \sim - i \frac{\sqrt{2}}{15} \frac{N^{\frac{5}{2}}}{\sqrt{8-N_f}} \frac{[(\Delta+\hat{m})(\Delta-\hat{m})]^{\frac{3}{2}}}{\omega_1 \omega_2}\ ,
\end{equation}
subject to the constraint
\begin{equation}
\Delta - \omega_1 -\omega_2 =  2\pi i\ ,
\end{equation}
in the Cardy-like limit $|\omega_i| \ll 1$.
$\Delta^2-\hat{m}^2$ appearing in the square-root is negative in our canonical chamber.
The expression (\ref{free-energy}) and similar expressions at the end of
section 2.2 are obtained with the convention $(-1)^{\frac{3}{2}}=-i$.
In section 2.3, we shall explain that this free energy counts the dual BPS black holes
in the background of warped AdS$_6\times S^4/\mathbb{Z}_2$ product. Here
we simply note that the leading large $N$ free energy $\propto N^{\frac{5}{2}}$ does not
see the flavor symmetries, e.g. $M_l$'s and $q$ for $SO(2N_f)\times U(1)_I
\subset E_{N_f+1}$.
This is natural in the bulk dual because the states charged under $E_{N_f+1}$ are
localized on a codimension $1$ wall, $S^3\sim\partial(S^4/\mathbb{Z}_2)$, so that the
leading large $N$ bulk physics does not see them. However, the value of $N_f$ itself is
visible in the leading free energy. This is because the number of D8-branes affects the
bulk dilaton field.

\subsection{$SU(2N)$ theories}\label{sec: Cardy-SU(N)}

Similar to the studies of section \ref{sec: Cardy-Sp(N)}, we analyze the large $N$ and
Cardy free energy of the index for 5d $\mathcal{N}=1$ gauge theory with
$SU(2N)$ gauge group, two rank 2 antisymmetric hypermultiplets, and $N_f$
fundamental hypermultiplets. The related geometric settings are explained at the
beginning of this section. The index is defined as \cite{Kim:2012gu}
\begin{equation}
\hspace*{-0.5cm}
Z(\omega_1, \omega_2, \Delta, m, M, q) = \textrm{Tr}\left[(-1)^F e^{-\beta\{Q,S\}} e^{-\omega_1 J_1} e^{-\omega_2 J_2} e^{-(\Delta-2\pi i) R} e^{-m h_M- b h_B}
e^{- \sum_{l} M_l H_l} q^k \right]\ .
\end{equation}
We introduced the fugacities $e^{-m}, \, e^{-b}$ for the Cartans
$h_M, \, h_B$ of $SU(2)_M \times U(1)_B \cong U(2)$ acting on two
antisymmetric matters, and the fugacities $e^{-M_l}$ for
the Cartans $H_l$ of $U(N_f)$ acting on fundamental matters.
As before, the parameters should meet the constraint
$\Delta-\omega_1-\omega_2=2\pi i$ for $Z$ to be an index.
Again, the index is given by \cite{Kim:2012gu, Bergman:2013koa}
\begin{eqnarray}
\hspace*{-0.5cm}
Z(\omega_{1,2}, \Delta, m, b, M_l, q) &=& \oint [d\alpha] \textrm{PE} \big
[f_{vec} (\omega_{1,2}, \Delta, \alpha_a) + f_{mat}^{asym} (\omega_{1,2}, m, b, \alpha_a)
+ f_{mat}^{fund} (\omega_{1,2}, M_l, \alpha_a)\big]\nonumber
\\
&&\times\prod_{\pm}Z_{\rm inst} (\pm\omega_1, \pm\omega_2, \pm\Delta, \pm m,
\pm b, \pm M_l, \pm\alpha_a,q^{\pm 1})\ .
\end{eqnarray}
The Haar measure is given by
$[d\alpha] = \frac{1}{(2N)!} \left[\prod_{a} \frac{d\alpha_a}{2\pi} \right] \prod_{a<b} \left[2 \sin \left(\frac{\alpha_a-\alpha_b}{2}\right) \right]^2$,
with $\sum_{a=1}^{2N} \alpha_a =0$.
Sum of all $\alpha_a$'s vanishes because the gauge group is
$SU(2N)$. The letter indices are given by
\begin{eqnarray}
\hspace*{-0.5cm}
\tilde{f}_{vec} &=& - \frac{2\cosh \frac{\tilde{\Delta}}{2}}{2\sinh \frac{\omega_1}{2}
\cdot 2\sinh \frac{\omega_2}{2}} \left[\sum_{a<b}^{2N} \left(e^{-i\alpha_a+i\alpha_b}+e^{i\alpha_a-i\alpha_b}\right) \right]
+ \left(\!1 - \frac{2\cosh \frac{\tilde{\Delta}}{2}}{2\sinh \frac{\omega_1}{2}
\cdot 2\sinh \frac{\omega_1}{2}} \!\right) (2N\!-\!1)\ , \nonumber\\
\hspace*{-0.5cm}
f_{mat}^{asym}&=& \frac{2 \cosh \frac{m}{2}}{2\sinh \frac{\omega_1}{2} \cdot 2\sinh \frac{\omega_2}{2}} \sum_{a<b}^{2N} \left(e^{\frac{b}{2}+i\alpha_a+i\alpha_b}+e^{-\frac{b}{2}-i\alpha_a-i\alpha_b} \right) \ , \nonumber\\
\hspace*{-0.5cm}
f_{mat}^{fund}&=& \frac{1}{2\sinh \frac{\omega_1}{2} \cdot 2\sinh \frac{\omega_2}{2}} \sum_{l=1}^{N_f} \sum_{a=1}^{2N} \left(e^{M_l+i\alpha_a}
+e^{-M_l-i\alpha_a} \right)\ ,
\end{eqnarray}
where $\tilde{\Delta}=\Delta-2\pi i$. As in section 2.1,
the Haar measure contribution is absorbed into
$\tilde{f}_{vec}$. The instanton part $Z_{\rm inst}$ can be
computed from the ADHM construction of multi-instantons \cite{Hwang:2014uwa}.
We assume $Z_{\rm inst}\approx 1$ at the large $N$ saddle point that we shall
present below. We believe this can be shown using the methods of \cite{Hwang:2014uwa}.
With this assumed, and following the steps similar to section \ref{sec: Cardy-Sp(N)},
PE of the letter indices in the Cardy limit $|\omega_i|\ll 1$ are given by
\begin{eqnarray}
  \textrm{PE}\left[\tilde{f}_{vec}\right] &\sim&
  \exp \Bigg[ - \frac{1}{\omega_1 \omega_2} \sum_{a<b}^{2N} \sum_{\pm,\pm} \textrm{Li}_3(-e^{\pm \frac{\Delta}{2}\pm i(\alpha_a-\alpha_b)}) \Bigg]
  \equiv \exp \left[ - \frac{\mathcal{F}_{vec}(\alpha_a, \, \Delta)}{\omega_1 \omega_2}\right]\ ,\nonumber\\
  \textrm{PE} \left[f_{mat}^{asym} \right] &\sim&
  \exp \Bigg[ \frac{1}{\omega_1 \omega_2} \sum_{a<b}^{2N} \sum_{\pm,\pm} \textrm{Li}_3(e^{\pm\frac{m}{2} \pm( \frac{b}{2}+i\alpha_a+i\alpha_b)}) \Bigg]
  \equiv \exp \left[ - \frac{\mathcal{F}_{mat}^{asym}(\alpha_a, \, m, \, b)}{\omega_1 \omega_2}\right]\ ,\nonumber\\
  \textrm{PE} \left[f_{mat}^{fund} \right] &\sim&
  \exp \Bigg[ \frac{1}{\omega_1 \omega_2} \sum_{l=1}^{N_f} \sum_{a=1}^{2N}\sum_{\pm} \textrm{Li}_3(e^{\pm(M_l+i\alpha_a)})\Bigg]
\equiv \exp \left[ - \frac{\mathcal{F}_{mat}^{fund}(\alpha_a, \, M_l)}{\omega_1 \omega_2}\right]\ .
\end{eqnarray}
The index is then given by the following expression,
\begin{equation}\label{Cardy-index-2}
\begin{aligned}
Z(\omega_1, \, \omega_2, \, \Delta, \, m, \, b, \, M_l) &\sim \frac{1}{(2N)!} \oint \prod_{a=1}^{2N} \frac{d\alpha_a}{2\pi} \exp \left[ - \frac{\mathcal{F}^{pert}(\alpha_a, \, \Delta, \, m, \, b, \, M_l)}{\omega_1 \omega_2}  \right] \\
&\equiv \frac{1}{(2N)!} \oint \prod_{a=1}^{2N} \frac{d\alpha_a}{2\pi} \exp \left[ - \frac{\mathcal{F}_{vec}+\mathcal{F}_{mat}^{asym}+\mathcal{F}_{mat}^{fund}}{\omega_1 \omega_2}  \right]\ .
\end{aligned}
\end{equation}

We study the saddle point in the large $N$ and Cardy limit.
We again take the following ansatz
for the eigenvalue distribution:
\begin{equation}\label{ansatz-2}
\alpha_a= i N^\alpha \tilde{x}_a\ , \quad \sum_{a=1}^{2N} \tilde{x}_a=0\ ,
\end{equation}
where $\tilde{x}_a$'s are of order $\mathcal{O}(N^0)$, and $0<\alpha<1$. We order $\tilde{x}_a$'s to be increasing using the Weyl symmetry of $SU(2N)$, i.e. $\tilde{x}_1<\tilde{x}_2<\cdots < \tilde{x}_{2N}$.
With the ansatz \eqref{ansatz-2}, $\mathcal{F}_{mat}^{asym}$ is given by
\begin{eqnarray}\label{b-asym}
\mathcal{F}_{mat}^{asym}&=& - \sum_{a<b}^{2N} \sum_\pm \Big[\textrm{Li}_3 (e^{-N^\alpha (\tilde{x}_a+\tilde{x}_b) \pm \frac{m}{2} + \frac{b}{2}}) + \textrm{Li}_3 (e^{N^\alpha (\tilde{x}_a+\tilde{x}_b) \pm \frac{m}{2}-\frac{b}{2}}) \Big] \nonumber\\
&\sim & \, \frac{1}{3} N^{\alpha} \sum_{a<b}^{2N} \bigg[\left\{\frac{3}{2}\frac{m+b}{2}\left(\frac{m+b}{2}-2\pi i\right) + \frac{3}{2}\frac{m-b}{2}\left(\frac{m-b}{2}-2\pi i\right)- 2\pi^2\right\}|\tilde{x}_a+\tilde{x}_b| \nonumber\\
&&\qquad \qquad \ \ \
- \frac{3}{2} N^{\alpha} b \; \textrm{sgn}(\tilde{x}_a+\tilde{x}_b) |\tilde{x}_a+\tilde{x}_b|^2 +N^{2\alpha} |\tilde{x}_a+\tilde{x}_b|^3 \bigg]\ ,
\end{eqnarray}
where $m\pm b$ are understood to be in the range $(0,4\pi i)$.
Here, we used the trilogarithm formulae \eqref{trilog-zero}, \eqref{trilog-large} at $N \to \infty$. Similarly, $\mathcal{F}_{mat}^{fund}$, $\mathcal{F}_{vec}$ are given by
\begin{equation}\label{pre-fund}
\mathcal{F}_{mat}^{fund} = - \sum_{l=1}^{N_f} \sum_{a=1}^{2N} \left[\textrm{Li}_3 (e^{-N^\alpha \tilde{x}_a + M_l}) + \textrm{Li}_3 (e^{N^\alpha \tilde{x}_a - M_l})\right] 
\sim \frac{1}{6} N^{3\alpha} \sum_{l=1}^{N_f} \sum_{a=1}^{2N} |\tilde{x}_a|^3 
= \frac{N_f}{6} N^{3\alpha} \sum_{a=1}^{2N} |\tilde{x}_a|^3 
\end{equation}
and
\begin{eqnarray}\label{pre-vec}
\mathcal{F}_{vec} &=& \sum_{a<b}^{2N} \sum_{\pm} \Big[\textrm{Li}_3 (-e^{-N^\alpha (-\tilde{x}_a+\tilde{x}_b)\pm \frac{\Delta}{2}}) + \textrm{Li}_3 (-e^{N^\alpha (-\tilde{x}_a+\tilde{x}_b)\pm \frac{\Delta}{2}}) \Big]
\sim  \sum_{a<b}^{2N} \sum_{\pm} \textrm{Li}_3 (e^{N^\alpha (-\tilde{x}_a+\tilde{x}_b)\pm \frac{\omega_1+\omega_2}{2}}) \nonumber\\
&\sim &- \frac{1}{3} N^{\alpha} \sum_{a<b}^{2N} \bigg[\left(3\left(\frac{\Delta}{2} + \pi i\right)\left(\frac{\Delta}{2} - \pi i\right) - 2\pi^2\right)(\tilde{x}_b-\tilde{x}_a) +N^{2\alpha} (\tilde{x}_b-\tilde{x}_a)^3 \bigg]\ ,
\end{eqnarray}
where $\Delta\approx 2\pi i$, and we used $\Delta-\omega_1-\omega_2 = 2\pi i$.
Again, we have shown apparently subleading terms in $N^{-1}$ in foresight,
which will turn out to be dominant after extremization and cancelations. Collecting all, $\mathcal{F}^{pert}=\mathcal{F}_{vec}+\mathcal{F}_{mat}^{asym}+\mathcal{F}_{mat}^{fund}$
is given by
\begin{eqnarray}\label{3alpha-prepot}
\mathcal{F}^{pert} &\sim& \frac{N_f}{6} N^{3\alpha} \sum_{a=1}^{2N} |\tilde{x}_a|^3
+ \frac{1}{2}N^{\alpha}\!\sum_{a\neq b}^{2N} \left(\gamma_m |\tilde{x}_b+\tilde{x}_a|\!-\!\gamma_\Delta |\tilde{x}_b-\tilde{x}_a|\right)  - \frac{\pi^2}{3}N^{\alpha}\!\sum_{a\neq b}^{2N} \left(|\tilde{x}_b+\tilde{x}_a|- |\tilde{x}_b-\tilde{x}_a|\right) \nonumber \\
&&-\frac{1}{4} N^{2\alpha} b \sum_{a \neq b}^{2N} \textrm{sgn}(\tilde{x}_a+\tilde{x}_b) |\tilde{x}_a+\tilde{x}_b|^2 + \frac{1}{6} N^{3\alpha} \sum_{a\neq b}^{2N} \left(|\tilde{x}_b+\tilde{x}_a|^3-|\tilde{x}_b-\tilde{x}_a|^3\right)\ , \nonumber\\
&&\gamma_m \equiv \, \frac{1}{2} \sum_\pm \frac{m \pm b}{2} \left(\frac{m \pm b}{2}-2\pi i\right)\ , \quad \gamma_\Delta \equiv \left(\frac{\Delta}{2} + \pi i\right)\left(\frac{\Delta}{2} - \pi i \right) \ .
\end{eqnarray}

At this moment, the leading contribution to \eqref{3alpha-prepot} at large $N$ comes
from the last term which is of order $O(N^{2+3\alpha})$. So we extremize the last term.
The analysis is similar to \cite{Jafferis:2012iv}.
To find a saddle point of the last term, we define
\begin{equation}
x_{N+1-a} \equiv \frac{\tilde{x}_{2N+1-a}-\tilde{x}_a}{2}\ , \quad y_{N+1-a} \equiv \frac{\tilde{x}_{2N+1-a}+\tilde{x}_a}{2} \quad (1\leq a \leq N)\ .
\end{equation}
Let us first consider the extremization with $y_{N+1-a}$'s. Differentiating the last
term of \eqref{3alpha-prepot}, one obtains
\begin{eqnarray}\label{vanish-eqn}
0&=&\frac{\partial}{\partial y_{N+1-i}} \sum_{a\neq b}^{2N} \left(|\tilde{x}_b+\tilde{x}_a|^3-|\tilde{x}_b-\tilde{x}_a|^3\right) = \left(\frac{\partial}{\partial \tilde{x}_{2N+1-i}} + \frac{\partial}{\partial \tilde{x}_{i}} \right) \sum_{a\neq b}^{2N} \left(|\tilde{x}_b+\tilde{x}_a|^3-|\tilde{x}_b-\tilde{x}_a|^3\right) \nonumber\\
&=&6 \sum_{a\neq i}^{2N} \left(\textrm{sgn}(\tilde{x}_i+\tilde{x}_a)(\tilde{x}_i+\tilde{x}_a)^2
-\textrm{sgn}(\tilde{x}_i-\tilde{x}_a)(\tilde{x}_i-\tilde{x}_a)^2\right) \\
&&+6 \sum_{a\neq 2N+1-i}^{2N} \left(\textrm{sgn}(\tilde{x}_{2N+1-i}+\tilde{x}_a)(\tilde{x}_{2N+1-i}+\tilde{x}_a)^2
-\textrm{sgn}(\tilde{x}_{2N+1-i}-\tilde{x}_a)(\tilde{x}_{2N+1-i}-\tilde{x}_a)^2\right)\ ,
\nonumber
\end{eqnarray}
where $1 \leq i \leq N$. One can find that a solution is given by \cite{Jafferis:2012iv}
\begin{equation}\label{vanish-con}
-\tilde{x}_a=\tilde{x}_{2N+1-a} = x_{N+1-a} \quad (1 \leq a \leq N)\ .
\end{equation}
So on this solution, we can take $N$ variables
$x_a$'s as the remaining variables to extremize with.
They are ordered as $0\leq x_1<x_2< \cdots<x_N $.
Note that this solution satisfies the condition $\sum_{a=1}^{2N} \tilde{x}_a=0$.
As in \cite{Jafferis:2012iv}, we assume that this solution for $\delta y_{N+1-a}$ variation
is the relevant one for our problem.
Then, the remaining problem is to extremize with $x_a$'s. Inserting the saddle point solution for $y_a$'s \eqref{vanish-con} to the last term of \eqref{3alpha-prepot},
one finds
\begin{equation}
\left.\frac{N^{3\alpha}}{6} \sum_{a\neq b}^{2N} \left(|\tilde{x}_b+\tilde{x}_a|^3-|\tilde{x}_b-\tilde{x}_a|^3\right) \right|_{-\tilde{x}_a=\tilde{x}_{2N+1-a}} = - \frac{8N^{3\alpha}}{3} \sum_{a=1}^{N} x_a^3 = O(N^{1+3\alpha})\ .
\end{equation}
This is of the same order as the first term of \eqref{3alpha-prepot}.
So from now on, one should also consider all other terms in  \eqref{3alpha-prepot}
at the same order.
The possible leading terms in \eqref{3alpha-prepot} are of order $O(N^{1+3 \alpha})$ and $O(N^{2+2\alpha})$ at large $N$. However, imposing \eqref{vanish-con}, one can easily
check that the terms at order $O(N^{2+2\alpha})$ vanish because of the $\textrm{sgn}(\tilde{x}_a+\tilde{x}_b)$ factor in those terms. Then we are finally left
with terms at $O(N^{1+3\alpha})$ and $O(N^{2+\alpha})$ orders. These two terms will be balanced and provide leading terms.
Note that there are also subleading terms at $O(N^{1+\alpha})$ order after inserting
\eqref{vanish-con} into \eqref{3alpha-prepot}, which we ignore.

So inserting \eqref{vanish-con} into \eqref{3alpha-prepot}, one obtains
\begin{eqnarray}\label{SU(N)-eff-pre}
\hspace*{-1cm}\mathcal{F}^{pert} &\!\sim\!& \, -\frac{8-N_f}{3} N^{3\alpha} \sum_{a=1}^{N} x_a^3 + \frac{\gamma}{2}N^{\alpha} \sum_{a\neq b}^{2N}  |\tilde{x}_b+\tilde{x}_a| \\
&\!\sim\!& \, -\frac{8-N_f}{3} N^{3\alpha} \sum_{a=1}^{N} x_a^3 + \gamma N^{\alpha} \sum_{a< b}^{N} \left( 2(x_b+x_a) + 2(x_b-x_a) \right)
= \, -\frac{8-N_f}{3} N^{3\alpha} \sum_{a=1}^{N} x_a^3 + 4\gamma N^{\alpha} \sum_{a< b}^{N} x_b
\nonumber\\
&\!\sim\! & \, -\frac{8-N_f}{3} N^{1+3\alpha} \int_0^{x_*} dx \rho(x) x^3 + 4\gamma N^{2+\alpha} \int_0^{x_*} dx \rho(x) \int_x^{x_*} dx' \rho(x') x'\ , \nonumber
\end{eqnarray}
where
\begin{equation}
\gamma \equiv \gamma_m - \gamma_\Delta = \frac{1}{2} \sum_\pm \frac{m \pm b}{2} \left(\frac{m \pm b}{2}-2\pi i \right)- \left(\frac{\Delta}{2} + \pi i\right)\left(\frac{\Delta}{2} - \pi i\right)>0
\end{equation}
if we take $\Delta\approx 2\pi i$ and $m\pm b$ purely imaginary within the canonical
range $(0,4\pi i)$. The above effective action is essentially the same as that for the $Sp(N)$ gauge theory, \eqref{eff-pre}. The only difference is that double integral part for the $SU(2N)$ gauge theory is twice that of $Sp(N)$ gauge theory. Thus, the remaining extremization procedures are the same as those presented in section 2.1.
The resulting free energy is given by
\begin{equation}\label{free-energy-Z2}
\log Z \sim - i \frac{\sqrt{2}}{15} \frac{N^{\frac{5}{2}}}{\sqrt{8-N_f}} \frac{\left[(\Delta+(\hat{m}+b))(\Delta-(\hat{m}+b))+(\Delta+(\hat{m}-b))(\Delta-(\hat{m}-b))\right]^{\frac{3}{2}}}{\omega_1 \omega_2}\ ,
\end{equation}
where $\Delta-\omega_1-\omega_2 =  2\pi i$, $\hat{m} \equiv m-2\pi i$.

It is straightforward to generalize this result to the quivers for the
general $\mathbb{Z}_{n}$ orbifold \cite{Bergman:2012kr,Bergman:2013koa}.
Assuming $Z_{\rm inst}\approx 1$, we simply present the final
result for the large $N$ and Cardy free energy:
\begin{eqnarray}\label{free-energy-Zn}
  \hspace*{-0.3cm}
  \log Z &\sim& - i \frac{\sqrt{2}}{15} \frac{N^{\frac{5}{2}}}{\sqrt{8-N_f}} \frac{\left[\sum_{A}(\Delta\!+\!(\hat{m}\!+\!b_A))(\Delta\!-\!(\hat{m}\!+\!b_A))
  +2\sum_{I}
  (\Delta\!+\!(\hat{m}\!+\!b_I))(\Delta\!-\!(\hat{m}\!+\!b_I))\right]^{\frac{3}{2}}}{\omega_1 \omega_2}\ , \nonumber\\
  &&N_f \equiv \sum_{q}N_f^{(q)} \leq 7\ ,\ \hat{m} \equiv m-2\pi i\ ,\
  \sum_A b_A + 2 \sum_I b_I = 0\ ,\ \Delta-\omega_1-\omega_2 =  2\pi i\ .
\end{eqnarray}

\subsection{AdS$_6$ black holes}\label{sec: BH}

In this subsection, we explain that the large $N$ Cardy free energies derived in
sections 2.1 and 2.2 account for the BPS black holes in the dual AdS$_6$ backgrounds.
A crucial ingredient is the universal entropy function of such black holes found in
\cite{Choi:2018fdc}.

In principle, general black holes in AdS$_6$ can carry various
electric charges dual to the R-charge, mesonic charge, and baryonic charges.
However, BPS black hole solution in AdS$_6$ known to date was found in
6d $\mathcal{N}=(1,0)$ $SU(2)$ gauged supergravity \cite{Chow:2008ip}. For instance,
this 6d theory can be obtained by a consistent Kaluza-Klein
truncation of massive type IIA supergravity on $S^4/\mathbb{Z}_2$ \cite{Cvetic:1999un}.
This black hole solution has only one electric charge, corresponding to the $SU(2)$
R-charge $R$. So to compare our field theory results with known AdS$_6$ black holes of
\cite{Chow:2008ip}, we should perform Legendre transformations of
the free energies \eqref{free-energy}, \eqref{free-energy-Z2}, \eqref{free-energy-Zn}
at zero mesonic and baryonic charges.
Firstly, at generic $n\geq 2$, one should extremize the following entropy function,
\begin{equation}\label{entropy-Zn}
S(\Delta, m, b_{A,I}, \omega_i;\; R, Q_M, Q_{B_{A,I}}, J_i)
= \log Z + \Delta R + m Q_M + \sum_{A,I} b_{A,I} Q_{B_{A,I}} + \omega_1 J_1 + \omega_2 J_2\ ,
\end{equation}
subject to the constraints
\begin{equation}\label{constraint-Zn}
\sum_A b_A + 2 \sum_I b_I =0\ , \quad \Delta-\omega_1-\omega_2=2\pi i\ .
\end{equation}
$\log Z$ is given by either (\ref{free-energy-Z2}) or \eqref{free-energy-Zn}.
To compare with known black holes, we set
\begin{equation}
Q_M=0\ , \quad Q_{B_{A}}=0\ ,\ \ Q_{B_I}=0\ .
\end{equation}
Let us first consider the baryonic chemical potentials. For $SU(2N)$ gauge theory
at $n=2$, one should extremize
\begin{equation}
- i \frac{\sqrt{2}}{15} \frac{N^{\frac{5}{2}}}{\sqrt{8-N_f}} \frac{\left[(\Delta+(\hat{m}+b))(\Delta-(\hat{m}+b))
+(\Delta+(\hat{m}-b))(\Delta-(\hat{m}-b))\right]^{\frac{3}{2}}}{\omega_1 \omega_2} + b \, Q_B
\end{equation}
with $b$ at $Q_B=0$. The extremal solutions are given by
$b=0$, $\pm\sqrt{\Delta^2-\hat{m}^2}$. However, for the latter two solutions, one finds
that $\log Z=0$ after inserting these values of $b$. Making further Legendre transformation
of $\log Z=0$ in $\omega_{1,2}$'s and $\hat{m}$ would yield zero entropy, which means
that $b=\pm\sqrt{\Delta^2-\hat{m}^2}$ will not yield the dominant saddle point.
So we take the solution $b=0$. Similarly, for the most general case with $n\geq 2$,
one can easily show that the dominant saddle point at zero baryon charges is given by
\begin{equation}\label{extreme-baryon}
b_{A}=0\ ,\ \ b_{I}=0\ .
\end{equation}
Inserting this solution \eqref{extreme-baryon} to \eqref{entropy-Zn}, one obtains
\begin{equation}
S(\Delta, m, \omega_i; R, Q_M, J_i) = - i \frac{\sqrt{2}}{15} \frac{n^{\frac{3}{2}}N^{\frac{5}{2}}}{\sqrt{8-N_f}} \frac{[(\Delta+\hat{m})(\Delta-\hat{m})]^{\frac{3}{2}}}{\omega_1 \omega_2} + \Delta R
+ m Q_M + \omega_1 J_1 + \omega_2 J_2\ ,
\end{equation}
where $\hat{m}=m-2\pi i $. Here,
note that for the $Sp(N)$ gauge theory at $n=1$, the free energy (\ref{free-energy})
of section 2.1 agrees with the above formula at $n=1$. So one can use this entropy
function for $\Delta,m,\omega_{1,2}$ as universally describing the 5d SCFTs labelled
by $n\geq 1$ at zero baryon charges.

We then Legendre transform in $m$, at $Q_M=0$.
The saddle points for $\hat{m}$ variation at $Q_M=0$ are given by
\begin{equation}
  \hat{m}=0\ ,\ \ \pm\Delta\ .
\end{equation}
Again, the latter two solutions have $\log Z=0$, so that further Legendre transformation
with $\omega_{1,2}$ will yield zero. So the dominant saddle point is given by
\begin{equation}
\hat{m}=0 \; \rightarrow \; m=2\pi i\ .
\end{equation}
Inserting this solution for $m$, we finally obtain the following entropy function:
\begin{equation}\label{micro-entropy}
S=- i \frac{\sqrt{2}}{15} \frac{n^{\frac{3}{2}}N^{\frac{5}{2}}}{\sqrt{8-N_f}} \frac{\Delta^{3}}{\omega_1 \omega_2} + \Delta R + \omega_1 J_1 + \omega_2 J_2\ ,
\end{equation}
subject to the constraint $\Delta - \omega_1 -\omega_2 =  2\pi i$.
This form of entropy function was shown in \cite{Choi:2018fdc} to precisely
account for the entropy and chemical potentials of BPS black holes in AdS$_6$. However,
the entropy function there was expressed universally, in terms of the Newton constant $G$ of
6d gauged supergravity instead of the microscopic parameters $n, N, N_f$ of our models.
In the remaining part of this subsection, we explain the conversion of these parameters to
establish the microscopic account for the BPS black holes.

To find the relation between $G$ and $N$, $N_f$, $n$,
we need the explicit metric of AdS$_6 \times (S^4/\mathbb{Z}_2)/\mathbb{Z}_n$ in massive type IIA supergravity. It is a warped product of AdS$_6$ with radius $\ell$ and half of $S_4/\mathbb{Z}_n$ with radius $\frac{2\ell}{3}$.
The 10d metric in string frame is given by \cite{Brandhuber:1999np}
\begin{equation}
ds_{10}^2 = \frac{1}{(\sin \alpha)^{\frac{1}{3}}} \left[\ell^2 ds^2(AdS_6)
+ \frac{4\ell^2}{9} \left(d\alpha^2 +\cos^2 \alpha \, ds^2(S^3/\mathbb{Z}_n)\right) \right]\ ,
\end{equation}
where $ds^2(AdS_6)$ is the metric of AdS$_6$ with unit radius,
and $ds^2(S^3/\mathbb{Z}_n)$ is the metric for $S^3/\mathbb{Z}_n$, whose volume is $\textrm{vol}(S^3/\mathbb{Z}_n)=\frac{2\pi^2}{n}$. The range of $\alpha$ is given by $\left(0, \frac{\pi}{2}\right]$. The gauge coupling constant $g$ in 6d gauged supergravity is related to the radius $\ell$ of AdS$_6$ by $g=\ell^{-1}$
\cite{Cvetic:1999un, Chow:2008ip}.
Also, from the quantization of the 4-form flux, $\ell$ is related
to $N$ by \cite{Jafferis:2012iv, Bergman:2012kr}
\begin{equation}
\frac{\ell^4}{\ell_s^4} = \frac{18\pi^2 nN}{8-N_f}\ ,
\end{equation}
where $\ell_s$ is the string scale.
We will also need the dilaton field, given by \cite{Jafferis:2012iv}
\begin{equation}
e^{-2\Phi} = \frac{3(8-N_f)^{\frac{3}{2}}\sqrt{nN}}{2\sqrt{2}\pi} (\sin \alpha)^{\frac{5}{3}}\ .
\end{equation}

The 10d Newton constant is given by $2 \kappa_{10}^2=16\pi G_{10}=(2\pi)^7 \ell_s^8$ \cite{Jafferis:2012iv}. The 6d Newton constant
is obtained by reducing the 10d Einstein-Hilbert action on $(S^4/\mathbb{Z}_2)/\mathbb{Z}_n$,
down to 6d Einstein-Hilbert action. During this reduction, the 6d metric $g_{\mu\nu}$ is
embedded into the 10d metric $G_{MN}$ as
\begin{equation}
  ds_{10}^2=G_{MN}dx^Mdx^N=\frac{1}{(\sin\alpha)^{\frac{1}{3}}}
  \left[g_{\mu\nu}dx^\mu dx^\nu+\frac{4\ell^2}{9}\left(d\alpha^2+\cos^2\alpha~
  ds^2(S^3/\mathbb{Z}_n)\right)\right]\ .
\end{equation}
The 10d Einstein-Hilbert action reduces to 6d as
\begin{equation}\label{10-6-metric}
  \frac{1}{G_{10}}\int d^{10}x\sqrt{-G}e^{-2\Phi}G^{MN}R_{MN}(G)\
  \longrightarrow\ \frac{1}{G}\int d^6x\sqrt{-g}g^{\mu\nu}R_{\mu\nu}(g)\ .
\end{equation}
This leads to the following relation:
\begin{eqnarray}\label{G-convert}
\frac{1}{G} &=& \frac{1}{G_{10}} \int_{(S^4/\mathbb{Z}_2)/\mathbb{Z}_n} d^4x
\sqrt{G/g}~ e^{-2\Phi} \times (\sin\alpha)^{\frac{1}{3}} \nonumber\\
&=&\frac{1}{2^3\pi^6 \ell_s^8}  \int_0^{\frac{\pi}{2}} d\alpha \; \sqrt{\left(\frac{4\ell^2}{9}\right)^4 (\cos^2 \alpha)^3} \cdot
\textrm{vol} (S^3/\mathbb{Z}_n) \cdot \frac{3(8-N_f)^{\frac{3}{2}}\sqrt{nN}}{2\sqrt{2}\pi}
(\sin \alpha)^{\frac{5}{3}} \cdot (\sin \alpha)^{-\frac{4}{3}} \nonumber\\
&
=&\frac{\sqrt{2} \ell^4}{3^3\pi^5 \ell_s^8} (8-N_f)^{\frac{3}{2}}\sqrt{\frac{N}{n}} \; \frac{9}{20}
=\frac{27\sqrt{2}}{5\pi\ell^4} \frac{n^{\frac{3}{2}}N^{\frac{5}{2}}}{\sqrt{8-N_f}}\ .
\end{eqnarray}
Here the factor $(\sin\alpha)^{\frac{1}{3}}$ on the first line comes from
the relative factor between $G^{MN}$ and $g^{\mu\nu}$ appearing in
(\ref{10-6-metric}). Using \eqref{G-convert}, \eqref{micro-entropy} can be
rewritten as
\begin{equation}\label{universal-entropy-ftn}
S=- i \frac{\pi}{81g^4G} \frac{\Delta^{3}}{\omega_1 \omega_2} + \Delta R
+ \omega_1 J_1 + \omega_2 J_2\ ,
\end{equation}
subject to the constraint $\Delta - \omega_1 -\omega_2 =  2\pi i$.
This in fact is the universal formula found in \cite{Choi:2018fdc} for any R-charged
BPS black holes in AdS$_6$, irrespective of its string theory embedding.
In \cite{Choi:2018fdc}, it has been shown that extremizing the above
entropy function, and imposing the characteristic charge relation \cite{Chow:2008ip}
satisfied by these black holes, one obtains the Bekenstein-Hawking entropy and chemical
potentials of such black holes.

Before briefly summarizing the key results of \cite{Choi:2018fdc}, let us
comment on the intrinsic studies that can be made from the index.
Since $\Delta=2\pi i+\omega_1+\omega_2$, $S$ takes the form of
\begin{equation}\label{entropy-ftn-rearrange}
  S=-\frac{\pi i}{81g^4G}\frac{(2\pi i+\omega_1+\omega_2)^3}{\omega_1\omega_2}
  +\omega_1(R+J_1)+\omega_2(R+J_2)+2\pi iR\ .
\end{equation}
Therefore, only the two combinations $R+J_1$, $R+J_2$ of charges appear nontrivially
in the Legendre transformation, which is natural since this is the free energy
of the index. The saddle point value $S_\ast(R,J_{1,2})$ after the extremization would
be complex. One should really consider the degeneracy rather than entropy, so
we study $e^{S_\ast}$. This takes the following form:
\begin{equation}
  e^{S_\ast(R,J_1,J_2)}=e^{2\pi iR+i{\rm Im}f(R+J_1,R+J_2)}\cdot
  e^{{\rm Re}f(R+J_1,R+J_2)}\ .
\end{equation}
Here, $f$ is a complex function of $R+J_1$, $R+J_2$ that one obtains
after extremizing the first three terms of (\ref{entropy-ftn-rearrange}).
The first factor is a phase factor which depends on the macroscopic charges
$R,J_1,J_2$, which rapidly oscillates as one changes these charges.
For instance, let us first consider the factor $e^{2\pi i R}$. Although $R$
is macroscopic, we know that $R$ is quantized to be a
half-integer. Then by changing $R$ by its minimal quantized unit,
$e^{2\pi iR}$ will hop between $+1$ and $-1$.
However, it looks highly unclear in general whether the whole phase factor
$e^{i(2\pi R+{\rm Im}f)}$ is real and hops between $\pm 1$ as the charges
are changed by quantized units. At the dominant saddle with largest
$e^{{\rm Re}f}$, one may change the logic here and demand that the unitarity of QFT
guarantees this phase factor to be either $\pm 1$. It appears meaningless to try to
check this with the results at hand. This is because we have made a macroscopic saddle
point approximation at large charges, and such quantized properties are generally
expected to be visible only after including subleading corrections.
Anyway, in this strategy, ${\rm Re}f(R+J_1,R+J_2)$ would be the macroscopic
entropy that one can extract out intrinsically from the index, dressed by
the $\pm 1$ factor which is represented by a phase in our macroscopic calculus.
This has been often the attitudes assumed in \cite{Choi:2018hmj,Choi:2018vbz}.

Now to summarize some key results of \cite{Choi:2018fdc}, we first note that
the known BPS black holes of \cite{Chow:2008ip} carry two independent parameters.
So the charges $R,J_1,J_2$ satisfy a relation upon the known solutions.
\cite{Choi:2018fdc} imposed this relation, and showed that
${\rm Re}f$ agrees with the Bekenstein-Hawking entropy $S_{\rm BH}$
of these black holes.\footnote{Technically, one finds
${\rm Im}S_\ast=2\pi R+{\rm Im}f=0$ after imposing the charge relation, so that
$f$ on these solutions is actually real and equals $S_{\rm BH}$. We
lack an intrinsic QFT understanding, if any, of this phenomenon.}
The resulting $S_{\rm BH}={\rm Re}f$ is given by \cite{Choi:2018fdc}
\begin{eqnarray}
  S_{\rm BH}^3-\frac{2\pi^2}{3g^4G}S_{\rm BH}^2-12\pi^2R^2S_{\rm BH}
  +\frac{8\pi^4}{3g^4G}J_1J_2&=&0\nonumber\\
  RS_{\rm BH}^2+\frac{2\pi^2}{9g^4G}(J_1+J_2)S_{\rm BH}-\frac{4\pi^2}{3}R^3&=&0\ .
\end{eqnarray}
This is a result derived from QFT by imposing extra charge relation by hand.
Solving these two equations, $S_{\rm BH}$ acquires two apparently different
expressions in terms of $R,J_1,J_2$. The compatibility of the two expressions
is the charge relation imposed. It was shown in \cite{Choi:2018fdc} that the
Bekenstein-Hawking entropy of the black holes of \cite{Chow:2008ip} satisfies
precisely the same equations. This establishes the QFT account for the
BPS black holes in AdS$_6$.

Since we have derived $\log Z$ of the dual SCFTs in the Cardy limit
$|\omega_{1,2}|\ll 1$, we have microscopically derived the thermodynamics of
corresponding large BPS black holes in AdS$_6$.
The Cardy limit $|\omega_{1,2}|\ll 1$ on the known black hole solutions demands
$J_1,J_2\gg R\gg N^{5/2}$. Similar to the AdS$_5$/CFT$_4$ models studied in the literature
\cite{Choi:2018hmj,Choi:2018vbz}, we generally expect that there could be more complicated
and yet unknown black hole saddle points beyond the Cardy limit. However, as
shown by \cite{Benini:2018ywd} in AdS$_5$/CFT$_4$, the known black
holes should still represent local large $N$ saddle points, irrespective of
whether they are most dominant or not. Here, we note that the entropy function
(\ref{universal-entropy-ftn}) was shown to describe known black holes even beyond
the Cardy limit \cite{Choi:2018fdc}. If the instanton corrections
$Z_{\rm inst}$ can still be ignored to $\approx 1$ in the large $N$ non-Cardy regime,
it may be technically doable to search for such saddle points. This is beyond
the scope of this paper.

\subsection{Comments on instantons and 5d deconfinement}

While making the saddle point approximations in sections 2.1 and 2.2, we
used the perturbative parts of the index only.
Here, one might feel confused about the following point.
From large $N$ perturbative Yang-Mills theory,
one would not expect more than $N^2$ degrees of freedom. One expects to capture
some interesting SCFT physics from formulae like \cite{Kim:2012gu} through the instanton
part $Z_{\rm inst}$ in the integrand. But in all the large $N$ analyses in the literature
for 5d SCFTs, it naively appears that only the perturbative integrand contributes to
the large $N$ free energy, with $Z_{\rm inst}\approx 1$ suppressed. So one may wonder
if there are any roles played by $Z_{\rm inst}$ at all. We would like to comment that
it plays a subtle role in `disallowing' the $N^2$ scaling of the free energy.

Let us first ask the following question. Had the integrand for the index only consisted of
the perturbative part,
\begin{equation}\label{pert-index}
  Z_{\rm pert}(x,y,\{m\})=\oint[d\alpha]{\rm PE}
  \left[f_{vec}(x,y,e^{i\alpha_a})+\sum_{\bf R}
  f_{mat}^{\bf R}(x,y,e^{i\alpha_a},e^{m_{\bf R}})\right]
\end{equation}
without factors like $Z_{\rm inst}$, what would have been the expected Cardy free energy
in the limit $|\omega_{1,2}|\ll 1$? (Here, ${\bf R}$ runs over representations of
the gauge group $G$ for hypermultiplets.) The natural answer is already
presented in \cite{Choi:2018hmj,Honda:2019cio,ardehali} for 4d
$\mathcal{N}=4$ gauge theory,
and is extended to 4d $\mathcal{N}=1$ theories in \cite{Kim:2019yrz}. Namely, in 4d SUSY gauge theories
whose indices take the form of (\ref{pert-index}), the Cardy saddle point for the
gauge holonomies $\alpha_a$ is such that $G$ is unbroken at the saddle point.
In other words, all $e^{i\alpha_a}$ appearing in
$f_{vec}$ and $f_{mat}^{\bf R}$ can be effectively set to $e^{i\alpha_a}=1$, so
that the system is maximally deconfining in the Cardy limit.\footnote{In 4d
Yang-Mills theories, this has been naturally assumed in the literature,
e.g. in \cite{Aharony:2003sx} inspired by the high temperature limit of the
Gross-Witten-Wadia model \cite{Gross:1980he,Wadia:1980cp}.}

Had the 5d index been just (\ref{pert-index}), we would naturally expect the same
holonomy saddle structure because the letter indices basically take
the same forms. Most importantly, the letter indices take the form of
$\frac{1}{(1-xy)(1-x/y)}=\frac{1}{(1-e^{-\omega_1})(1-e^{-\omega_2})}$ times finite
polynomials of fugacities, both in 4d and 5d. So it is natural to expect the same
large $N$ and Cardy saddle point structures for these integrals. At these saddle points,
the free energy of the index (\ref{pert-index}) would be proportional to
$N^2$, naturally agreeing with the combinatoric interpretation of this formula which
counts gauge invariant operators of the free theory.
Therefore, although $Z_{\rm inst}$ can be ignored at the final stage of our saddle point
analyses in sections 2.1 and 2.2,  $Z_{\rm inst}$ should somehow
play subtle intermediate roles to disfavor the
saddle point $e^{i\alpha_a}\sim 1$, rather preferring the complexified saddle point with
$-i\alpha_a\sim N^{\frac{1}{2}}\gg 1$.

In fact, one can see that the possibility of the saddle point $e^{i\alpha_a}\sim 1$
becomes highly unclear with the presence of $Z_{\rm inst}$,
for the following reason. For instance, 
the $1$ instanton part of $Z_{\rm inst}$ in our $Sp(N)$ theory
is given by \cite{Kim:2012gu,Hwang:2014uwa}
\begin{eqnarray}
  Z_1&=&\frac{1}{2}\left[\frac{\prod_{l=1}^{N_f}2\sinh\frac{M_l}{2}\prod_{a=1}^N
  2\sinh\frac{m\pm i\alpha_a}{2}}
  {2\sinh\frac{\omega_{1,2}}{2}\cdot
  2\sinh\frac{m\pm\omega_+}{2}\prod_{a=1}^N2\sinh\frac{\omega_+\pm i\alpha_a}{2}}
  +\frac{\prod_{l=1}^{N_f}2\cosh\frac{M_l}{2}\prod_{a=1}^N
  2\cosh\frac{m\pm i\alpha_a}{2}}
  {2\sinh\frac{\omega_{1,2}}{2}\cdot
  2\sinh\frac{m\pm\omega_+}{2}\prod_{a=1}^N2\cosh\frac{\omega_+\pm i\alpha_a}{2}}\right]
  \nonumber\\
  &&-\frac{1}{2}\frac{\prod_{l=1}^{N_f}2\sinh\frac{M_l}{2}+\prod_{l=1}^{N_f}2\cosh\frac{M_l}{2}}
  {2\sinh\frac{\omega_{1,2}}{2}\cdot 2\sinh\frac{m\pm\omega_+}{2}}\ ,
\end{eqnarray}
where $\omega_+\equiv\frac{\omega_1+\omega_2}{2}$. 
In the Cardy limit $|\omega_{1,2}|\ll 1$, this becomes
\begin{equation}
  Z_1\sim\frac{1}{2\omega_1\omega_2}\left[
  \frac{\prod_{l=1}^{N_f}2\sinh\frac{M_l}{2}}{2\sinh\frac{m}{2}}
  \left(\prod_{a=1}^N\frac{2\sinh\frac{m\pm i\alpha_a}{2}}{2\sinh\frac{\pm i\alpha_a}{2}}-1\right)
  +\frac{\prod_{l=1}^{N_f}2\cosh\frac{M_l}{2}}{2\sinh\frac{m}{2}}
  \left(\prod_{a=1}^N\frac{2\cosh\frac{m\pm i\alpha_a}{2}}{2\cosh\frac{\pm i\alpha_a}{2}}-1\right)
  \right]\ .
\end{equation}
This diverges at $\alpha_a=0$ (and also at $\alpha_a=\pi$). From the physics of instantons,
this divergence is due to the non-compact zero mode of instanton size becoming massless.
More physically, if one expands $Z_1$ in the fugacities
$e^{i\alpha_a}$'s in the Coulomb branch, at ${\rm Im}(\alpha_a)>0$,
one finds infinite towers of BPS bound states
with increasing $U(1)^N\subset Sp(N)$ electric charges,
since $2\sinh(\frac{i\alpha_a}{2})$ factors appear in the denominator. So the
divergence at $\alpha_a=0$ comes from these infinite towers of non-perturbative
charged states
in the Coulomb branch, if one ceases to weight them by fugacity factors $e^{i\alpha_a}$.
Since this divergence is caused by replacing $\sinh\frac{\omega_+\pm i\alpha_a}{2}$ by
$\sinh\frac{\pm i\alpha_a}{2}$, the divergence actually represents an extra factor of
$\frac{1}{(2\sinh\frac{\omega_+}{2})^{2N}}\sim\frac{1}{\omega_+^{2N}}$ in the
naive Cardy limit. As one goes to higher instanton numbers $k>1$, there
appear more infinite towers of charged fields. The extra divergent factor
becomes $\frac{1}{\omega^{2Nk}}$. An easy way to see this is to note that there are
$2c_2k=2(N+1)k$ complex zero modes in the $k$ instanton background, where
$c_2=N+1$ is the dual Coxeter number of $Sp(N)$. Among these, $2k$ comes from the
position zero modes of $k$ instantons, so that it only causes
$\frac{1}{\omega_1\omega_2}$ divergence in the free energy.
The remaining $2Nk$ complex zero modes come from the internal
degrees of freedom, yielding extra $\frac{1}{\omega^{2Nk}}$ factor.

Had this been the true saddle point, the Cardy free energy is not behaving like
the one for a reasonable 5d CFT, which we expect to be proportional to
$\frac{1}{\omega_1\omega_2}$ times a coefficient representing the number of
degrees of freedom in this CFT (which is $\sim N^{\frac{5}{2}}$ in our problem).
It is not even clear whether the sum over $k$ would make sense.\footnote{There
is even a signal that this is a divergent series \cite{Marino:2015nla}.
We thank Antonio Sciarappa for telling this to us.}
So collecting all, we find that it is highly unclear whether
$e^{i\alpha_a}\sim 1$ is a legitimate saddle point
in the presence of the $Z_{\rm inst}$ factor.

On the other hand, as we have seen in sections 2.1 and appendix B,
the saddle point with
$-i\alpha_a\sim\mathcal{O}(N^{\frac{1}{2}})$ has suppressed $Z_{\rm inst}\approx 1$,
and one can self-consistently show that only the `perturbative integrand' needs to
be considered.\footnote{It may be misleading to simply call it
`perturbative' part, which often refers to the perturbative non-Abelian
gauge theory. It should be more precisely stated as the 1-loop Coulomb branch contribution,
with both massive instantons and W-bosons integrated out (whose masses are proportional
to large Coulomb VEV).} As a result of such spreading of eigenvalues,
it apparently seems that $N^2\cdot N^{\frac{1}{2}}\sim N^{\frac{5}{2}}$ enhancement
happened, if we just consider it in the context of the partition function (\ref{pert-index}).
However, with $Z_{\rm inst}$ factor in mind, we think this interpretation is
misleading. This is because the instanton part of the free energy at $e^{i\alpha_a}\sim 1$
rapidly grows as $\sim \omega^{-2Nk}$ in $k$, possibly reflecting
an inconsistency of the (grand) canonical ensemble due to the rapid growth
of density of states as $k$ is increased. Compared to this, the saddle point
with $-i\alpha_a\sim\mathcal{O}(N^{\frac{1}{2}})$ exhibits a sensible growth of
free energy in $\omega^{-1}$. The former is perhaps analogous to the Hagedorn-like growth of density of states in the confining phase of 4d free QFT \cite{Aharony:2003sx}, which is
made much more extreme in 5d by the additional infinite towers of instanton bound states.

Deconfinement in AdS$_5$/CFT$_4$ implies that the growth of density of states
is slowed down after the transition. This is made possible by breaking the infinite
towers of `hadrons' into deconfined quark-gluon
partons. It is also associated with absorbing latent heat during the transition,
after which extensive quantities show the enhancement $N^0\rightarrow N^2$ in large $N$.
From the gauge theory side, this is
achieved by setting $e^{i\alpha_a}$'s closer to $1$. In AdS$_6$/CFT$_5$, it seems that there
should be more ingredients to achieve the exotic deconfinement in 5d SCFTs. Taking
$e^{i\alpha_a}\sim 1$ partly liberates quarks and gluons, from the viewpoint of
non-renormalizable perturbative gauge theories. However, the system still has infinite
towers of bound states made with instanton solitons. A wild speculation that has been
made in the literature is that these instantons are also made of certain partons
\cite{Collie:2009iz,Bolognesi:2011nh}. The non-compact
internal zero modes were interpreted as the position moduli of such hypothetical partons.
If such a conjecture is true, liberating the instanton partons
should make $N^2\rightarrow N^{\frac{5}{2}}$ enhancement possible, while the
rapidly growing density of states in the `energy' $k$ will be tamed after this
deconfinement. From the gauge theory viewpoint, understanding the distinctions between
the 4d saddles $e^{i\alpha_a}\sim 1$ and our 5d saddles $-i\alpha_a\sim N^{\frac{1}{2}}$
should encode some details of such hypothetical deconfinement. In a sense, one can regard
the real $i\alpha_a$'s as the `inverse-temperature'
variables for the electric charges. The saddle point $i\alpha_a=0$ of the
4d Cardy free energy can be understood as such `temperatures' sent to infinity, to
maximally liberate the quark-gluon partons. In our 5d gauge theory analysis,
the true saddle point with $i\alpha_a<0$ may be understood as going 
beyond this infinite temperature point. This looks like a natural direction 
in which the remaining instanton-partons can be liberated.
We would very much like to make such speculations more precise in the future.

Supplementing the thoughts in the previous paragraph, we finish this subsection by
contrasting the differences between apparently similar 4d and 5d indices. Namely,
with the nonperturbative completion of (\ref{pert-index}) by $Z_{\rm inst}$, we
argued that the saddle point with $e^{i\alpha_a}\sim 1$ is obstructed (or at least
its existence is made non-obvious) by nontrivial $Z_{\rm inst}$. For certain 5d
gauge theories, we instead explored alternative large $N$ saddle points in which
$-i\alpha_a\sim N^{\frac{1}{2}}$. To make the speculations of previous paragraph
more sensible, one would like to make an obvious sanity check that similar
saddle points with analytically continued $\alpha_a$ do not exist in 4d indices,
which also take the form of (\ref{pert-index}). In particular, dimensionally reducing
our main 5d examples given by $Sp(N)$ gauge theories, one obtains
4d $\mathcal{N}=2$ gauge theories. Since the 4d system
makes sense as long as $N_f\leq 4$ with very similar field contents,
one may technically wonder if similar analytically continued saddle points
can exist in its 4d version. However, we have explicitly checked that such
analytically continued saddle points do not exist in the 4d $\mathcal{N}=2$ index,
even for precisely the same gauge theories reduced to 4d. Therefore, the analytically
continued saddle point which is in charge of $N^{\frac{5}{2}}$ scaling is indeed
a 5d phenomenon.

\section{Conclusions}

In this paper, we studied the index of a class of 5d SCFTs on $S^4\times\mathbb{R}$,
by taking the large $N$ and Cardy limit. Our large $N$ Cardy free energy precisely
accounts for the thermodynamic properties of large BPS black holes in global AdS$_6$.

The basic calculus is very similar to those made in different supersymmetric partition
functions \cite{Jafferis:2012iv,Hosseini:2018uzp,Crichigno:2018adf}. In our context, like
\cite{Hosseini:2018uzp,Crichigno:2018adf}, the gauge holonomies $e^{i\alpha_a}$ have to be analytically
continued away from the unit circle to reach the relevant saddle point. At the final
stage of calculus, the so-called instanton correction to the partition function is
suppressed to $Z_{\rm inst}\approx 1$ at our saddle point.
We have discussed the physical meanings of such a calculus, pointing out the subtle roles
of the instanton part and contrasting it to the indices of 4d QFTs.
This has close relations to the mysterious deconfinements in 5d SCFTs. Our results
should shed concrete lights on getting a better physical picture of such
novel deconfinements, and hopefully a better quantitative picture on the instanton
partons.

We have focussed on a very small subset of 5d SCFTs, engineered on
D4-branes in massive type IIA string theory. Recently, a much broader
class of 5d SCFTs have been discovered: e.g. see \cite{Jefferson:2017ahm,Jefferson:2018irk}
for geometric engineering, and \cite{Hayashi:2018bkd} for
brane engineering. Also, there have been explorations on the
large $N$ AdS$_6$ duals of 5d SCFTs, engineered by the 5-brane webs
\cite{DHoker:2016ujz,Apruzzi:2014qva,Kim:2015hya}. In the generic setting in which
the numbers of external $(p,q)$ 5-branes are at similar order $\sim N$,
various physical quantities are known to scale like $N^4$ \cite{DHoker:2016ujz}.
Although we find these examples more difficult to study in our framework,
perhaps numerical studies similar to those of \cite{Fluder:2018chf,Fluder:2019szh}
could be made.

\vskip 0.5cm

\hspace*{-0.8cm} {\bf\large Acknowledgements}
\vskip 0.2cm

\hspace*{-0.75cm} We thank Hee-Cheol Kim, Joonho Kim, Sung-Soo Kim, Kimyeong Lee and
Antonio Sciarappa for helpful discussions. This work is supported in part by the National
Research Foundation of Korea (NRF) Grant 2018R1A2B6004914 (SC, SK), NRF-2017-Global
Ph.D. Fellowship Program (SC), a KIAS Individual Grant PG081602 at Korea Institute for Advanced Study (SC), and the NRF grant 2021R1A2C2012350 (SK).

\appendix

\section{Trilogarithm function}\label{appen: trilog}

The trilogarithm function is defined by a power series in $z$ when $|z|<1$ as
\begin{equation}\label{power-trilog}
\textrm{Li}_3 (z) = \sum_{n=1}^\infty \frac{z^n}{n^3} \quad (|z|<1)\ ,
\end{equation}
which can be extended to $|z| \geq 1$ by the process of analytic continuation. Note that the trilogarithm function is multi-valued. It has a branch point at $z=1$, and we take the principal branch with a branch cut along $(1,  +\infty)$.

When the argument goes to zero, the trilogarithm function goes to zero as
\begin{equation}\label{trilog-zero}
\lim_{x \to \infty} \textrm{Li}_3 (e^{-x+iy}) \approx e^{-x+iy} \; \to \; 0\ .
\end{equation}

The trilogarithm function satisfies the following inversion formula:
\begin{equation}\label{trilog-inv}
\begin{aligned}
\textrm{Li}_3 (e^a) - \textrm{Li}_3 (e^{-a}) &= - \frac{(2\pi i )^3}{3!} B_3 \left(\frac{a}{2\pi i} -p \right) \quad \left(2\pi p < \textrm{Im} (a) < 2\pi (p+1)\right) \\
&=-\frac{(a-2\pi i p)^3}{6} + \frac{\pi i (a-2\pi i p)^2}{2} + \frac{\pi^2 (a-2\pi i p)}{3}\ ,
\end{aligned}
\end{equation}
where $B_3 (a)$ is the third Bernoulli polynomial and $p$ is an integer. Then, using \eqref{trilog-zero}, \eqref{trilog-inv}, one obtains the following asymptotic formula
\begin{equation}\label{trilog-large}
\lim_{x \to \infty} \textrm{Li}_3 (e^{x+iy}) \sim -\frac{(x+iy-2\pi i p)^3}{6} + \frac{\pi i (x+iy-2\pi i p)^2}{2} + \frac{\pi^2 (x+iy-2\pi i p)}{3}
\end{equation}
in the range $2\pi p < y < 2\pi (p+1)$.

\section{Suppressed instantons in the Coulomb branch}

We explicitly evaluate the instanton partition function of 5d $\mathcal{N}=1$ $Sp(N)$ gauge theory, with 1 antisymmetric hypermultiplet and $N_f \leq 7$ fundamental hypermultiplets.
In particular, we estimate the suppression factor at large Coulomb VEV.
In this appendix, we take $\phi_a\equiv-i\alpha_a$ to be real, positive and large.
The instanton partition function is computed using the formulae and
prescriptions of \cite{Hwang:2014uwa}.

We start from the simpler case with $k=1$. The partition
function is given by \cite{Hwang:2014uwa}
\begin{eqnarray}
  Z_1&=&\frac{1}{2} \frac{1}{2\sinh \frac{\epsilon_+ \pm \epsilon_-}{2} 2\sinh \frac{m \pm \epsilon_+}{2}} \left[\prod_{l=1}^{N_f} 2\sinh \frac{M_l}{2} \left(\prod_{a=1}^N
  \frac{2\sinh \frac{m\pm \phi_a}{2}}{2\sinh \frac{\epsilon_+\pm \phi_a}{2}} -1 \right)
  \right.\\
  &&\left.+\prod_{l=1}^{N_f} 2\cosh \frac{M_l}{2} \left( \prod_{a=1}^N \frac{2\cosh \frac{m\pm \phi_a}{2}}{2\cosh \frac{\epsilon_+\pm \phi_a}{2}}-1 \right) \right]
  = \left\{
  \begin{array}{ll}
    \sim O(v_a) , & N_f \neq 0 \\
    \sim O(v_a^2), & N_f = 0
  \end{array}
  \right.\ , \nonumber
\end{eqnarray}
where $v_a \equiv e^{-\phi_a}\ll 1$ with $\phi_a\gg 1$.
We can interpret the above result as the contribution from the D0-brane stuck to O8-orientifold. One might naively think that, if D0-branes are bound to the O8-plane,
it will belong to the spectrum of string theory but not within the 5d QFT living on
the D4-branes. However, in the above expression for $Z_1$, such spurious states
coming from the D0-D8-O8 bound states are already eliminated, as found in
\cite{Hwang:2014uwa}. Although D0-branes are not bound to specific D4-branes,
they carry $O(v_a)$ or $O(v_a^2)$ suppression factors from their quantum masses
which depend on the Coulomb VEV. $O(v_a^2)$ contribution at $N_f=0$ carries one more
factor of $v_a$ since the periods of $\alpha_a$'s are doubled in the absence of
fundamental matters.

For $k \geq 2$, one should integrate over the $O(k)$ holonomy of the ADHM quantum mechanics
for $Sp(N)$ instantons, as explained in \cite{Hwang:2014uwa}. This yields a residue sum.
The residues can be classified according to the positions of D0-branes. When the pole
locations for the $O(k)$ holonomies are independent of the Coulomb VEV $\alpha_a$'s,
we interpret their residues as the contribution from the D0-branes stuck to O8.
When the $O(k)$ poles depend on $\alpha_a$'s, we interpret their residues as the
contribution from bound states of D0 and $a$'th D4-branes.

We first present the results for the $Sp(2)$ theory at $k=2,3,4$,
when $N_f \geq 1$. For $k=2$,
\begin{equation}
\begin{aligned}
&Z_2(0,0) \sim O(v_2^2, v_2 v_1, v_1^2)\ , \\
&Z_2(\phi_1, -\phi_1) \sim O(v_1^{8-N_f}, v_2^{-1}v_1^{9-N_f}, v_2^{-2} v_1^{10-N_f}, \ldots)\ , \\
&Z_2(\phi_2, -\phi_2) \sim O(v_2^{8-N_f}, v_2^{7-N_f} v_1, v_2^{6-N_f} v_1^2, \ldots)\ .
\end{aligned}
\end{equation}
Here, $0$ or $\pm \phi_a$'s are the positions of the D0-branes in
$\phi$ direction. $0$ indicates that the D0-brane is bound to O8, and
$\pm \phi_a$ indicate that it is bound to $a$'th D4-brane. For $k=3$,
\begin{equation}
\begin{aligned}
&Z_3(0,0,0) \sim O(v_2^3, v_2^2 v_1, v_2v_1^2, v_1^3)\ , \\
&Z_3(0,\phi_1, -\phi_1) \sim O(v_2 v_1^{8-N_f}, v_1^{9-N_f}, v_2^{-1} v_1^{10-N_f}, \ldots)\ , \\
&Z_3(0,\phi_2, -\phi_2) \sim O(v_2^{9-N_f}, v_2^{8-N_f} v_1, v_2^{7-N_f} v_1^2, \ldots)\ .
\end{aligned}
\end{equation}
For $k=4$,
\begin{equation}
\begin{aligned}
&Z_4(0,0,0,0) \sim O(v_2^4, v_2^3 v_1, v_2^2v_1^2, v_2v_1^3, v_1^4)\ , \\
&Z_4(0,0,\phi_1, -\phi_1) \sim O(v_2^2 v_1^{8-N_f}, v_2 v_1^{9-N_f}, v_1^{10-N_f}, \ldots)\ , \\
&Z_4(0,0,\phi_2, -\phi_2) \sim O(v_2^{10-N_f}, v_2^{9-N_f} v_1, v_2^{8-N_f} v_1^2, \ldots)\ , \\
&Z_4(\phi_1, -\phi_1,\phi_1, -\phi_1) \sim O(v_1^{2(8-N_f)}, v_2^{-1}v_1^{2(8-N_f)+1} , \ldots)\ , \\
&Z_4(\phi_1, -\phi_1,\phi_2, -\phi_2) \sim O(v_2^{8-N_f} v_1^{8-N_f}, v_2^{7-N_f} v_1^{9-N_f}, \ldots)\ , \\
&Z_4(\phi_2, -\phi_2,\phi_2, -\phi_2) \sim O(v_2^{2(8-N_f)}, v_2^{2(8-N_f)-1} v_1, \ldots)\ .
\end{aligned}
\end{equation}
We checked these formulae at $N_f=1$ for $k=1,2,3,4$, and at $N_f=2$ for $k=1,2$.
Therefore, even in the sector in which D0-branes are bound to D4-branes, their
quantum mass formula is somewhat complicated. However, if we take all $v_a$'s at
similar order, each D0-brane bound to a D4-brane carries a factor
$v^{8-N_f}$, consistent with the picture of \cite{Jafferis:2012iv}.
However, for D0's bound to O8, one finds a factor of $v_a\ll 1$ per D0-brane, while
the naive picture of \cite{Jafferis:2012iv} would have yielded $1$.

When $N_f=0$, the results are slightly different as follows.
\begin{equation}
\begin{aligned}
&Z_2(0,0) \sim O(v_2^4, v_2^3 v_1,v_2^2 v_1^2,v_2 v_1^3, v_1^4)\ , \\
&Z_2(\phi_1, -\phi_1) \sim O(v_1^{8}, v_2^{-1}v_1^{9}, v_2^{-2} v_1^{10}, \ldots)\ , \\
&Z_2(\phi_2, -\phi_2) \sim O(v_2^{8}, v_2^{7} v_1, v_2^{6} v_1^2, \ldots)\ , \\
&Z_3(0,0,0) \sim O(v_2^6, v_2^5 v_1, v_2^4 v_1^2, v_2^3 v_1^3, v_2^2 v_1^4, v_2 v_1^5,  v_1^6)\ , \\
&Z_3(0,\phi_1, -\phi_1) \sim O(v_2^2 v_1^{8}, v_2 v_1^{9}, v_1^{10}, \ldots)\ , \\
&Z_3(0,\phi_2, -\phi_2) \sim O(v_2^{10}, v_2^{9} v_1, v_2^{8} v_1^2, \ldots)\ .
\end{aligned}
\end{equation}

Summarizing the above results, we find that D0-D4 bound states
contribute as $O(v^{8-N_f})$ to the index (if we set $v_a$'s to be of similar order),
as we expected. However, there is another contribution to the index: D0-branes bound to O8-plane. When $N_f \geq 1$, they contribute as $O(v)$, while at $N_f=0$, they contribute as $O(v^2)$. For most values of $N_f$'s, the latter contribution is larger than the former.
In any case, we find $Z_k \sim \mathcal{O}(e^{-k\phi})$ or 
$\sim \mathcal{O}(e^{-2k\phi})$, meaning
that the instanton contribution is suppressed at large Coulomb VEV.

\end{document}